\documentclass[journal]{IEEEtran}
\usepackage{graphicx}
\usepackage{epstopdf}
\usepackage{float}
\usepackage{algorithmic}
\usepackage{algorithm, algorithmic}
\usepackage{array}
\usepackage{amsmath}
\usepackage{amssymb}
\usepackage{amsthm}
\usepackage{eqparbox}
\usepackage{fixltx2e}
\usepackage{xcolor}
\usepackage{makecell}
\usepackage{mathrsfs}
\usepackage{amsmath}
\usepackage{multicol}

\usepackage{url}
\usepackage{cite}
\usepackage[justification=centering]{caption}
\usepackage{bm}
\usepackage{threeparttable}
\usepackage{subeqnarray}
\usepackage{cases}
\ifCLASSOPTIONcompsoc
\usepackage[caption=false,font=normalsize,labelfont=sf,textfont=sf]{subfig}
\else
\usepackage[caption=false,font=footnotesize]{subfig}
\fi

\allowdisplaybreaks[4]

\newtheorem{mypro}{Proposition}

\newcommand{\diag}{\mathop{\mathrm{diag}}}



\begin{document}
\captionsetup{font={small}}		
	
	\title{Cooperative Beamforming Design for Multiple RIS-Assisted Communication Systems}
	
	\author{Xiaoyan Ma,~\IEEEmembership{Student Member, IEEE,}
		Yuguang Fang,~\IEEEmembership{Fellow, IEEE,} 
		Haixia~Zhang,\\\IEEEmembership{Senior Member, IEEE,}
		Shuaishuai~Guo,~\IEEEmembership{Member, IEEE,}
		and  Dongfeng~Yuan,~\IEEEmembership{Senior Member, IEEE}
	\thanks{This work was supported in part by the
		Project of International Cooperation and Exchanges NSFC under Grant No. 61860206005, in part by the National Natural Science Foundation of China under Grant No. 62171262 and in part by the 
		Major Scientific and Technological Innovation Project of Shandong Province
		2020CXGC010108 (\emph{Corresponding author: Haixia Zhang}).
	}
	
	\thanks{X. Ma and D. Yuan are with Shandong Key Laboratory of Wireless Communication Technologies, Jinan 250061, China, and School of Information Science and Engineering, Shandong University, Qingdao 266237, China (email: maxiaoyan06@mail.sdu.edu.cn; dfyuan@sdu.edu.cn).}
	\thanks{H. Zhang and S. Guo are  with Shandong Key Laboratory of Wireless Communication Technologies, Jinan 250061, China, and School of Control Science and Engineering, Shandong University, Jinan 250061, China (email: shuaishuai\textunderscore guo@sdu.edu.cn; haixia.zhang@sdu.edu.cn).}
	\thanks{Y. Fang is  with the Computer, Electrical and
		Mathematical Science and Engineering Division,  Department of Electrical and Computer Engineering
		University of Florida, Gainesville, FL, USA (email: fang@ece.ufl.edu).}	
	}	
\markboth{IEEE TRANSACTIONS ON WIRELESS COMMUNICATIONS,~Vol.~xx, No.~xx, 2022}%
{Shell \MakeLowercase{\textit{et al.}}: Bare Demo of IEEEtran.cls for IEEE Journals}

	\maketitle
	\IEEEpeerreviewmaketitle
	\begin{abstract}
		Reconfigurable intelligent surface (RIS) provides a promising way to build programmable wireless transmission environments. Owing to the massive number of controllable reflecting elements on the surface, RIS is capable of providing considerable passive beamforming gains. At present, most related works mainly consider the modeling, design, performance analysis and optimization of single-RIS-assisted systems.
		Although there are a few of works that investigate multiple RISs individually serving their associated users, the cooperation among multiple RISs is not well considered as yet. 
		To fill the gap, this paper studies a cooperative beamforming design for multi-RIS-assisted communication systems, where multiple RISs are deployed to assist the downlink communications from a base station to its users.
        To do so, we first model the general channel from the base station to the users for arbitrary number of reflection links. Then, we formulate an optimization problem to maximize the sum rate of all users.
        Analysis shows that the formulated problem is difficult to solve due to its non-convexity and the interactions among the decision variables.
        To solve it  effectively, we first decouple the problem into three disjoint subproblems.
Then, by introducing appropriate auxiliary variables, we derive the closed-form expressions for the decision variables and propose a low-complexity cooperative beamforming algorithm. 
		 Simulation results have verified the effectiveness of the proposed algorithm through comparison with various baseline methods. Furthermore, these results also unveil that, for the sum rate maximization, distributing the reflecting elements among multiple RISs is superior to deploying them at one single RIS.
	
	\end{abstract}	
	
	
	\section{Introduction}
	
	\IEEEPARstart{R}{econfigurable} intelligent surface (RIS) has emerged as a promising tool in improving the transmission environments for the wireless communications \cite{1,2,3}. Specifically, RIS can be a planar surface, which comprises a large number of low-cost passive reflecting elements. Each element can be digitally controlled and induce an independent amplitude change and/or phase shift to the incident signals, thereby collaboratively altering the wireless channels \cite{4,5,6}. Besides, from the implementation perspective, RISs possess appealing features such as low profile and lightweight, thus can be easily mounted on walls or ceilings without significantly damaging the aesthetic beauty of the environment   \cite{easydeploy}. Since RISs are complementary devices, deploying them in the existing wireless systems only need necessary modifications of the communication protocols, and no other changes on standards and hardware are required
	\cite{survey2}. 
	Furthermore, RISs are usually much cheaper than active small base stations (SBSs)/relays, therefore can be easily deployed rapidly \cite{cheaper2}.
	
	Due to such great advantages, RISs have attracted intensive attentions lately and a lot of related research works have been done, including transmission protocol designs \cite{protocol1,protocol2,protocol3}, system capacity analysis \cite{capacity1,capacity2,ZSW2,Ye1,SWZ1,Song}, energy/spectral efficiency \cite{efficiency1,efficiency2,efficiency3,efficiency4,efficiency5,reviewer32},
	physical layer security \cite{security1,security2,security3,security-multiple}, modulation/coding schemes \cite{coding1,coding2,coding3}, and so on. And a comprehensive introduction to RIS can be found in \cite{reviewer31}.
	Existing works mainly consider one or multiple distributed RISs, where each RIS individually reflects the signals to users, and there is no signal cooperation among multiple RISs. 
	This simplified approach generally contribute to sub-optimal performance \cite{Double2}, since the multi-hop links that can provide high cooperation beamforming gains are not well utilized.
    In addition, multi-hop reflections established through inter-RIS links can provide higher channel diversity, therefore can bypass the dense obstacles and provide blockage-free links with higher transmission quality. 
    
Inspired by the advantages of cooperative transmission in multi-RIS-assisted systems, Han {\sl et al.} investigated a double-RIS aided single-user wireless communication system. In \cite{MRIS1}, it is demonstrated that by cooperatively designing the phase shifts, a double-reflection link can provide passive beamforming gain that increases quartically with the total number of RISs' reflecting elements, which is significantly higher than the quadratic growth of the passive beamforming gain of single-RIS link with the same number of total reflecting elements.
This work has been further extended to multi-user setups \cite{Double2,MRIS2}, where the superiority of deploying double-RIS is further confirmed compared with single-RIS models. Specifically,
Zheng, You and Zhang \cite{Double2} have investigated a double-RIS-assisted multi-user uplink communication system. By jointly designing the receive beamforming at the BS and the cooperative reflection at these two RISs, the minimum signal-to-interference-plus-noise ratio (SINR) is maximized. 
With similar system setups made in \cite{Double2}, a double-RIS-assisted system is designed to enhance the secrecy performance \cite{MRIS2}. The above works only consider double RISs, Huang {\sl et. al} \cite{Huang} then extended to more general multi-RIS scenarios. In \cite{Huang}, only the multi-hop link established through multiple RISs is utilized to bypass the severe obstacles between the transmitter and the receivers, and their results also prove that the multiplicative beamforming gain obtained through inter-RIS channels can significantly improve the system capacity.

\begin{figure}[t]
	\centering
	\includegraphics[width=0.43\textwidth]{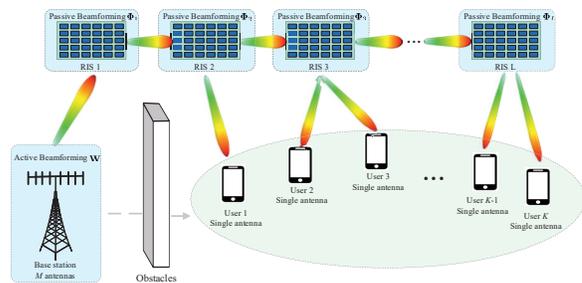}\\
	\caption{The proposed multi-RIS-assisted cooperative transmission system, where users at different locations can access the BS through proper RIS-assisted links.}
	\label{fig:system_model}
\end{figure}

Except the aforementioned works, general multi-RIS-assisted multi-user communication systems have not been fully investigated. Under the general setups with more available RISs, users at different locations can be served via various rich  paths established through multi-hop reflections as shown in Fig. \ref{fig:system_model}. 
And this gives rise to a new cooperative beamforming design problem, where multiple RISs serving multi-user through cooperative transmissions. In this work, we have been devoted to solve this problem to maximize the system sum rate. The main contributions are summarized as follows.
\begin{itemize}

\item We establish a multi-RIS-assisted multi-user communication system, where the general channel model for arbitrary reflection links is formulated. Different from the traditional multiple RIS-assisted systems, where only one multi-hop link that established through all the RISs is utilized. We address the possibility of distributed RISs in providing higher access flexibility. With this general setups, users at different locations can be served via various rich  paths established through multi-hop reflections, which significantly improve the system capacity.

\item Based on the proposed channel model, an optimization problem is formulated to maximize the sum rate of all users, which is multivariate and non-convex. To solve it effectively, we decompose the original problem into three disjoint sub-problems by utilizing closed-form fractional programming techniques. Specifically, we first decouple the sum-of-logarithms-of-ratio problem into sum-of-logarithms and sum-of-ratio problems by introducing auxiliary variables, then utilize quadratic transform to further simplify the sum-of-ratio part.

\item The formulated problem is a computationally intensive task due to the large number of RISs' reflecting elements and the interaction among different RISs. To facilitate the cooperative beamforming design, we reformulate the user's equivalent channel, so that the coupled phase shift matrix of each RIS can be extracted and iteratively updated. With the reformulated equivalent channel, we are also able to derive the closed-form optimal expression for each RISs' phase shift matrix, thereby significantly reduce the computational complexity.

\item To further simplify the overall computational complexity for cooperative beamforming design, we derive the closed-form optimal expressions for all the decision variables, and design a low-complexity cooperative beamforming algorithm to maximize the sum rate of the system. Analysis shows that the complexity of the proposed algorithm linearly increases with the number of RISs' reflecting elements. Extensive simulations have been conducted and demonstrated that the proposed algorithm achieves significantly higher sum rate than the baseline schemes. Furthermore, the results also unveil that distributing the reflecting elements among multiple RISs is preferable to deploying them at one single RIS in achieving higher sum rate.

\end{itemize}

	The remainder of this paper is organized as follows. The system model and problem formulation are introduced in Section II. In Section III, we introduce the proposed cooperative beamforming algorithm for multi-RIS-assisted multi-user communication systems. 
	To verify the superiority of the proposed method, simulation results are presented in Section V. Finally, we conclude this paper in Section VI.
	
	The notations used in this paper are listed as follows.
	Bold symbols in capital letter and small letter denote matrices and vectors, respectively.
	$\mathcal{CN}(\mu,\sigma^{2})$ denotes the circularly symmetric complex Gaussian (CSCG) distribution with mean $\mu$ and variance $\sigma^{2}$.
	$||\mathbf{w}||$ denotes the Euclidean norm while  $||\mathbf{W}||_{F}$ denotes the Frobenius norm.
	$\mathbf{I}_{M}$ represents the $M \times M$ identity matrix. $\mathbf{G}^{T}$ and $\mathbf{G}^{H}$ denote the transpose and conjugate transpose of matrix $\mathbf{G}$, respectively.  
	$\mathbf{G}^{-1}$ represents the  inverse of $\mathbf{G}$.
	$\otimes$ is the Kronecker product.
	$\mathbb{E}\lbrace \cdot\rbrace$ denotes the statistical expectation.                                                  
	For any real number, $\sqrt{a}$ denotes the square root of $a$.
	For any complex variable $x$, $|x|$ denotes the absolute value, $\angle x$ represents its phase angle, $x^{*}$ denotes its conjugate,  $\Re \lbrace x \rbrace$ and $\Im \lbrace x \rbrace$ represent its real part and imaginary part, respectively.
	$A_{i,j}$ represents the element at row $i$, column $j$ of matrix $\mathbf{A} $ and $a_{n}$ represents the $n$-th elements of vector $\bm{a}$. 
	$\mathbf{A}=\diag(\bm{a})$ means that $\mathbf{A}$ is the diagonal matrix of the vector $\bm{a}$. $\mathbf{0}$ is the zero matrix. 
	
	\section{System Model and Problem Formulation}
\subsection{System and Channel Model}
We consider a cooperative multi-RIS-assisted multi-user communication system as shown in Fig. \ref{fig:system_model}, in which $L$ distributed RISs are deployed to assist the downlink transmission from the $M$ antenna BS to $K$ single-antenna users. It is assumed that the direct links between the BS and users are severely blocked by the obstacles, and $L$ RISs are deployed between the BS and user cluster to assist the downlink transmissions. 
Without loss of generality, RIS $1$ is deployed very close to the BS to guarantee line-of-sight (LoS) channel, i.e., directly above the BS, other RISs (i.e., RIS $2,3,...,L$) are deployed sequentially to form multi-hop links.
The $l$-th RIS, $l=1,2,...,L$, is composed of $N_l$ reflecting elements, and each RIS is connected to a controller that adjusts its phase shifts. To be specific, 	
we consider the general case where there exist arbitrary reflection links in the communication systems. We let $\mathbf{G}_1 \in \mathbb{C}^{N_1 \times M}$, $\mathbf{H}_{l-1,l}\in\mathbb{C}^{N_l \times N_{l-1}}$ and $\bm{g}_{l,k}\in \mathbb{C}^{1 \times N_l}$ denote the channels from the BS to RIS $1$, from RIS $l-1$ to RIS $l$ and from RIS $l$ to user $k$, respectively. 
The reflection coefficient matrix of RIS $l$ can be expressed as
$\mathbf{\Phi}_{l} \in \mathbb{C}^{N_l \times N_l} $, which can be further expressed as $\mathbf{\Phi}_{l}= \diag (\pmb{\phi}_{l})$. $\pmb{\phi}_l=[\phi_{l,1},\phi_{l,2},...,\phi_{l,N_{l}}]^T$ is the corresponding reflection vector of the $l$-th RIS, where
$\phi_{l,n}=\beta_{l,n}e^{j\theta_{l,n}}$, $n = 1,2,...,N_{l}$. Moreover, $\theta_{l,n} \in [0,2\pi)$ and $\beta_{l,n} \in [0,1]$ represent the phase and amplitude change brought by the $n$-th reflecting element of RIS $l$ to the incident signals.
Without loss of generality, we set $\beta_{l,n} =1$, which implies that all the elements of the RISs are switched on to fully reflect the incident signals. 
%
%
%
%

\begin{figure}[t]
		\centering
		\includegraphics[width=0.4\textwidth]{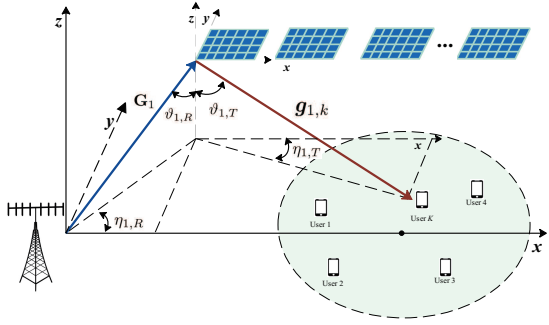}\\
		\caption{The channel model for multi-RIS-assisted system.}
		\label{fig:channel_model}
	\end{figure}

We assume that all the channels follow Rician fading. Similar to \cite{channel_model}, we further assume that the antenna elements form a half-wavelength uniform linear array (ULA) configuration at the BS, and the reflecting elements form a uniform planar array (UPA) configuration at RISs. RIS $l$ consists of a sub-wavelength UPA with $N_l= N_{lx}N_{ly}$ passive reflecting elements, where $N_{lx}$ and $N_{ly}$ denote the number of the reflecting elements along the x- and y-axis as shown in Fig. \ref{fig:channel_model}, the adjacent elements are separated by $d_x$ and $d_y$, respectively. We use $\mathbf{G}_1$ and $\bm{g}_{1,k}$ as examples to illustrate the adopted channel model in detail. Specifically, $\mathbf{G}_{1}$ can be modeled as
	\begin{equation} \label{G1}
		\mathbf{G}_1\!= \!\sqrt{\beta_{\mathbf{G}_1}} \left(\sqrt{\frac{\mathcal{F}_{\mathbf{G}_1}}{\mathcal{F}_{\mathbf{G}_1}+1}}\mathbf{G}_{1,LoS} \! +\! \sqrt{\frac{1}{\mathcal{F}_{\mathbf{G}_1}+1}} \mathbf{G}_{1,NLoS}\right),
	\end{equation}
	where $\beta_{\mathbf{G}_1}$ denotes the free-space path loss; $\mathcal{F}_{\mathbf{G}_1} $ represents the corresponding Rician factor; the matrices $\mathbf{G}_{1,LoS}$ and $\mathbf{G}_{1,NLoS}$ represent the line-of-sight (LoS) and non-LoS (NLoS) components of the channel $\mathbf{G}_1$, respectively.
	The LoS component $\mathbf{G}_{1,LoS}$ can be expressed as
	\begin{equation}
		\begin{aligned} \label{LoS_G1}
			\mathbf{G}_{1,LoS}= \mathbf{a}_{1,R}(\eta_{1,R},\vartheta_{1,R})\mathbf{a}_T^H(\vartheta_{1,R}),
		\end{aligned}
	\end{equation}
	where $\mathbf{a}_{1,R}(\eta_{1,R},\vartheta_{1,R}) \in \mathbb{C}^{N_1 \times  1}$ represents the receiving array response of RIS 1 and $\mathbf{a}_T(\vartheta_{1,R})\in \mathbb{C}^{M \times 1}$ denotes the transmit array response of the BS. 
	Let $\lambda$ denote the signal wavelength, by defining $\bar{d}_x = d_x/\lambda$ and $\bar{d}_y = d_y/\lambda$, we express the response vector $\mathbf{a}_{1,R}(\eta_{1,R},\vartheta_{1,R})$ as
	\begin{equation}
		\begin{aligned} 
			\mathbf{a}_{1,R}(\eta_{1,R},\vartheta_{1,R})=&
			[1,...,e^{j2\pi(N_{1x}-1)\bar{d}_x\Theta_{1,R}}]^T \\ &\otimes
			 [1,...,e^{j2\pi(N_{1y}-1)\bar{d}_y\Omega_{1,R}}]^T,
		\end{aligned}
	\end{equation}
	where $\Theta_{1,R}=\sin(\vartheta_{1,R})\cos(\eta_{1,R})$ and $\Omega_{1,R}=\sin(\vartheta_{1,R})\sin(\eta_{1,R})$ are the angle parameters with $\vartheta_{1,R}$ representing the zenith angle of arrival (AoA) and $\eta_{1,R}$ being the azimuth AoA at RIS 1 as shown in Fig. \ref{fig:channel_model}.
	The transmit array response  $\mathbf{a}_T(\vartheta_{1,R})$ can be similarly obtained as $\mathbf{a}_T(\vartheta_{1,R})=[1, e^{j\pi\sin \vartheta_{1,R}},...,e^{j\pi(M-1)\sin \vartheta_{1,R}}]^T$
	with $\vartheta_{1,R}$ representing the angle of departure (AoD) from the BS to RIS 1. The matrix
	$\mathbf{G}_{1,NLoS}$ represents the NLoS part, whose elements are chosen from $ \mathcal{CN}(0,1)$.
	Following the same processes, we obtain the channel $\bm{g}_{1,k}$ from RIS 1 to user $k$ as 
		\begin{equation}
		\begin{aligned}  \label{g1k}
		\bm{g}_{1,k}\!\!=\!\! \sqrt{\beta_{\bm{g}_{1,k}}}\!\! \left(\sqrt{\frac{\mathcal{F}_{\bm{g}_{1,k}}}{\mathcal{F}_{\bm{g}_{1,k}}+1}}\bm{g}_{1,k,LoS} \!\! +\!\! \sqrt{\frac{1}{\mathcal{F}_{\bm{g}_{1,k}}+1}} \bm{g}_{1,k,NLoS}\right),
		\end{aligned}
      \end{equation}
	where $\beta_{\bm{g}_{1,k}}$ represents the free-space path loss; $\mathcal{F}_{\bm{g}_{1,k}}$ is the corresponding Rician factor;  the vectors $\bm{g}_{1,k,LoS}$ and $\bm{g}_{1,k,NLoS}$ are the LoS and NLoS components of the channel, respectively. 
	The LoS component $\bm{g}_{1,k,LoS}$ can be represented by $\bm{g}_{1,k,LoS}=\mathbf{a}_{1,T}^T(\eta_{1,T},\vartheta_{1,T})$ with $\mathbf{a}_{1,T}(\eta_{1,T},\vartheta_{1,T})$ denoting the transmit array response of RIS 1. Similarly, by defining the angle parameters $\Theta_{1,T}=\sin(\vartheta_{1,T})\cos(\eta_{1,T})$ and $\Omega_{1,T}=\sin(\vartheta_{1,T})\sin(\eta_{1,T})$ with AoDs $\vartheta_{1,T}$ and $\eta_{1,T}$, we express
	$\mathbf{a}_{1,T}(\eta_{1,T},\vartheta_{1,T})$ as
	\begin{equation}
		\begin{aligned} 
			\mathbf{a}_{1,T}(\eta_{1,T},\vartheta_{1,T})=&
			[1,...,e^{j2\pi(N_{1x}-1)\bar{d}_x\Theta_{1,T}}]^T \\&  \otimes 
			[1,...,e^{j2\pi(N_{1y}-1)\bar{d}_y\Omega_{1,T}}]^T.
		\end{aligned}
	\end{equation}
	The vector $\bm{g}_{1,k,NLoS}$ in \eqref{g1k} is the NLoS component, whose elements are chosen from $ \mathcal{CN}(0,1)$. 
	The other channels can also be generated by following the same processes, we will not go into details here. The channel state information (CSI) for the BS-RIS $1$ link and the inter-RIS links can be considered to be known in advance, since once they are properly deployed, high-quality LoS channels can be established between them and those CSI can be directly utilized.
For the RIS-user channel information, there are two types of approaches to acquire CSI in the RIS-assisted systems, depending on whether the RIS's reflecting elements have the capability of sensing or not. 
For the case that RIS has only passive elements, since the RISs are usually properly deployed at the places that can establish LoS links to the users, such as the facade of the tall buildings, the location based channel estimation schemes can be used to get the CSI to assist the beamforming design \cite{location1,locationinfor2}. Besides, some other advanced channel estimation methods were also proposed to perform separate channel estimation in RIS-assisted systems. For example, Wei {\sl et al} \cite{seperate_CE} proposed a parallel factor decomposition based method to unfold the cascaded channel model, which can also be adopted. 
For the second case, where there is semi-passive RIS architecture, a small portion of the RIS's reflecting elements are able to process the received signal for facilitating the channel estimation. Based on the pilot signals received by these semi-passive elements, the RIS can estimate the channels between itself and the user through various existing estimation methods \cite{Semi-passive}.
With the CSI, the cooperative beamforming can be done. In the following, we will focus on the cooperative beamforming design for the multi-RIS-assisted communication system.
	\subsection{Problem Formulation}
	Let $x_k$ denote the transmitted signal from the BS to user $k$ with $\mathbb{E}\lbrace |x_k|^2  \rbrace=1, k=1,2,\cdots, K$.
	Then, the transmitted signals for all users at the BS can be expressed as
	\begin{eqnarray}
		\begin{aligned}
			\mathbf{x}=\sum_{k=1}^{K}\bm{ w}_{k}x_{k},
		\end{aligned}
	\end{eqnarray}
	where $\bm{w}_{k} \in \mathbb{C}^{M \times 1}$ is the corresponding transmit beamforming vector for user $k$. The equivalent channel from the BS to user $k$ can be represented as 
	\begin{eqnarray}
\begin{aligned} \label{Equ_channel}
\bm{h}_k^H=\bm{g}_{1,k}\mathbf{\Phi}_{1}\mathbf{G}_1
+\sum_{l=2}^{L}\bm{g}_{l,k}
\!\!\!\!\!\!
\prod_{i=l,l-1,...,2}\!\!\!\!\!\!\!\!\!\!\mathbf{\Phi}_{i}\mathbf{H}_{i-1,i} \mathbf{\Phi}_1\mathbf{G}_1,
\end{aligned}
\end{eqnarray} 
the first term $\bm{g}_{1,k}\mathbf{\Phi}_{1}\mathbf{G}_1$ in \eqref{Equ_channel} represents the single-reflection channel from the RIS $1$ to the user $k$,
the second term in \eqref{Equ_channel} is the multi-hop links established through the multiple RISs. Considering that the obstacle may very close to the BS, which can block the direct link from BS to the users and the other RISs excerpt for RIS $1$. Thereby in the proposed channel model, we let BS generate narrow beam with the total transmit power directly towards RIS $1$.
With the equivalent channel $\bm{h}_k^H$, the received signal at user $k$ can be expressed as
	\begin{eqnarray}
		\begin{aligned} \label{RS}
			y_{k}= \bm{h}_k^H\mathbf{x} + n_{k},
		\end{aligned}
	\end{eqnarray}
	where $n_{k} \sim \mathcal{CN}(0,\sigma_{0}^{2})$ denotes the complex additive white Gaussian noise (AWGN) at the $k$-th user. 
	The $k$-th user treats all the signals for other users as interference. Thus, the corresponding SINR at user $k$ can be expressed as
	\begin{equation}
\begin{split}\label{SINR_Ori}
			&\gamma_{k}=\frac{|\bm{h}_k^H \bm{w}_{k}|^2}
			{\sum_{i=1,i\neq k}^{K}
			|\bm{h}_k^H\bm{w}_{i}|^2+\sigma_{0}^{2}},
		\end{split}
	\end{equation}
	with
	$\sum_{k=1}^{K}||\bm{w}_k||^2\leq P_{T}$
	being the power constraint and $P_{T}$ is the maximum transmit power at the BS. When the BS utilizes all the power to transmit signals for user $k$, i.e., $||\bm{w}_k||^2=P_T$, the maximum SINR $\gamma_k^{max}=P_T||\bm{h}_k^H||^2 /\sigma_0^2$ can be achieved, therefore we have $0 \leq \gamma_k \leq \gamma_k^{max}$.
	
	Let $\mathbf{W}=[\bm{w}_{1},\bm{w}_{2},...,\bm{w}_{K}]\in \mathbb{C}^{M \times K}$ denote the overall transmit beamforming matrix. In this paper, we aim to maximize the sum rate of all users at the downlink transmission by jointly designing the active beamforming $\mathbf{W}$ at the BS and the passive beamforming $\pmb{\phi}_l$, $l=1,2,...,L$ at each RIS, subject to the BS's transmit power constraint and RISs' phase shift constraints.
	Mathematically, the optimization problem can be formulated as
	\begin{align} 
		(P1) \quad &\max_{\mathbf{W},\pmb{\phi}_1,...,\pmb{\phi}_L}  \quad f(\mathbf{W},\pmb{\phi}_1,...,\pmb{\phi}_L)=\sum_{k=1}^{K} \log(1+\gamma_{k}),\label{OBM}\\
		&\quad \quad \text{s.t.}\quad \quad |\phi_{l,n}|=1, ~ l=1,...,L,~ n=1,...,N_l, \label{PSC}\\
		& \quad\quad \quad \quad \quad \sum_{k=1}^{K}||\bm{w}_k||^2\leq P_{T}. \label{PC}
	\end{align}

		In ($P1$), the objective variables that need to be obtained are $\mathbf{W}$ and $\pmb{\phi}_1,...,\pmb{\phi}_L$, which are the active beamforming at the base station and the passive beamforming for RIS $l$, $l=1,2,...,L$. By cooperatively designing the transmit beamforming matrix $\mathbf{W}$ and the passive beamforming vectors $\pmb{\phi}_l$, the inter-user interference at the receiving sides can be mitigated, leading to the improvement of the sum rate. 
	However, solving ($P1$) is extremely challenging
	since the objective function $f(\mathbf{W},\pmb{\phi}_1,...,\pmb{\phi}_L)$ and the phase shift constraint \eqref{PSC} are non-convex. Moreover, the optimization variables $\mathbf{W}$ and $\pmb{\phi}_l$ are coupled with each other in the expression of the sum rate. All these bring great difficulties to find the global optimal solution. 
	 As an alternative, we attempt to find a low-complexity sub-optimal solution to ($P1$).

	\section{Low Complexity Beamforming Design for Multi-RIS-Assisted Multi-user Systems}
		In this section, the cooperative beamforming at both the BS and RISs are carefully designed to mitigate the  inter-user interference and maximize the sum rate. 
	In the original objective function \eqref{OBM}, $\sum_{k=1}^{K} \log(1+\gamma_{k})$ is a sum-of-log-of-ratio term, and the objective variables $\mathbf{W}$ and $\pmb{\phi}_l$ are coupled together.
	In order to solve the optimization problem efficiently, we first introduce several  auxiliary variables to transform ($P1$) into a problem of much low-complex by applying the closed-form fractional programming (FP) techniques \cite{FP1,FP2}. The closed-form FP techniques is proposed to deal with the sum-of-log-of-ratio problem as follows
	$$
	\max_{\bm{x}}\sum_{k=1}^{K}\log\left(1+\frac{|A_k(\bm{x})|^2}{B_k(\bm{x})-|A_k(\bm{x})|^2}\right),
	$$
	where $B_k(\bm{x})\geq|A_k(\bm{x})|^2$ for all $k$. Conceptually, the closed-form FP approach has two key steps:
	
	\emph{a) Lagrangian Dual Transform.} 
	By introducing an auxiliary variable $\alpha_k$, the logarithm function can be tackled based on the following equation
	$$
	\log(1+\gamma_k)=\max_{\alpha_k \geq 0} \left( \log(1+\alpha_k)-\alpha_k+\frac{(1+\alpha_k)\gamma_k}{1+\gamma_k}\right).
	$$
	And the equivalency is guaranteed when $\alpha_k=\gamma_k$. Then, the original problem is equivalently transformed to 	
	\begin{align}
		&\max_{\bm{x},\bm{\alpha}} \quad \sum_{k=1}^K \left(
		\log(1+\alpha_k)-\alpha_k+(1+\alpha_k)\frac{|A_k(\bm{x})|^2}{B_k(\bm{x})} \nonumber
		\right),\\
		&~\text{s.t.} \quad \quad \alpha_k \geq 0, \quad \forall \, k=1,2,...,K, \nonumber
	\end{align}
	where $\bm{\alpha}=[\alpha_1,\alpha_2,...,\alpha_K]^T$ is the auxiliary variable vector.
	
	\emph{b) Quadratic Transform.} Given $\bm{\alpha}$, we then focus on the following sum-of-ratios problem
	$$
	\max_{\bm{x}} \sum_{k=1}^{K}\frac{|A_k(\bm{x})|^2}{B_k(\bm{x})}.
	$$
	The key idea is introducing auxiliary variables $\bm{\beta}=[\beta_1,\beta_2,...,\beta_k]^T$, and then the above problem is equivalently translated to
	$$
	\max_{\bm{x},\bm{\beta}} \sum_{k=1}^{K} \left(
	2\Re\lbrace \beta_k^*A_k(\bm{x})\rbrace-|\beta_k|^2B_k(\bm{x})
	\right).
	$$
	The equivalence can be verified by substituting $\beta_k=\frac{A_k(\bm{x})}{B_k(\bm{x})}$ into the above equation.
	
Based on the reformulated problem, we aim to derive the closed-form optimal solutions for each decision variable. Specifically, the reformulated problem can be written as
	\begin{eqnarray} \label{ORP}
		\begin{aligned}
			(P2)  &\max_{\pmb{\alpha},\mathbf{W},\pmb{\phi}_1,...,\pmb{\phi}_L} \quad f_{2}(\pmb{\alpha},\mathbf{W},\pmb{\phi}_1,...,\pmb{\phi}_L),\\
			& \quad\quad\text{s.t.}\quad\quad\quad |\phi_{l,n}|\!\!=\!\!1, ~l=1,2,...,L, \\
			& \quad\quad\quad\quad\quad\quad\quad\quad\quad\quad~ n=1,2,...,N_l,\\
			& \quad\quad \quad\quad\quad\quad \sum_{k=1}^{K}||\bm{w}_k||^2\leq P_{T},\\
			&\quad \quad\quad \quad\quad\quad \alpha_{k}\geq 0, k=1,2,...,K,
		\end{aligned}
	\end{eqnarray}
	where
	\begin{eqnarray}
		\begin{aligned}\label{OBM1}
			f_{2}(\pmb{\alpha},\mathbf{W},\pmb{\phi}_1,...,\pmb{\phi}_L)
			=&\sum_{k=1}^{K} \log(1+\alpha_{k})-\sum_{k=1}^{K}\alpha_{k}\\
			&+\sum_{k=1}^{K}\frac{(1+\alpha_{k}) \gamma_{k}}{1+\gamma_{k}}.
		\end{aligned}
	\end{eqnarray}
	In $(P2)$, the objective variables that need to obtain are $\pmb{\alpha}$, $\mathbf{W}$ and $\pmb{\phi}_1,...,\pmb{\phi}_L$. $\pmb{\alpha}=[\alpha_{1},\alpha_{2},...,\alpha_{K}]^{T}$ is the auxiliary variable vector, each element $\alpha_k$ in it is an auxiliary variable that introduced for user $k$.
	 By doing so, we decouple the sum-of-log-of-ratio term $\sum_{k=1}^{K} \log(1+\gamma_{k})$ into the sum-of-log term $\sum_{k=1}^{K}\log(1+\alpha_{k})$ and the sum-of-ratio term $\sum_{k=1}^{K}\frac{(1+\alpha_{k}) \gamma_{k}}{1+\gamma_{k}}$ as shown in \eqref{OBM1}, which greatly reduces the difficulty in finding the optimal solutions. In the following, we further decouple $(P2)$ into three disjoint subproblems, and target to find the closed-form optimal solutions for each subproblem. 
	
	\emph{1) The closed-form optimal solution for $\pmb{\alpha}$.} 
	First, we aim to find the optimal closed-form expression of $\pmb{\alpha}$ for given $\mathbf{W}$ and $\pmb{\phi}_l$, $l=1,2,...,L$. By assuming the other parameters are fixed, the objective function in $(P2)$ can be seen as an optimization problem only with respect to $\pmb{\alpha}$, and we have the following proposition regarding its solution.
	\begin{mypro} \label{alpha1}
		When $\mathbf{W}$ and $\pmb{\phi}_1$, ..., $\pmb{\phi}_L$ are fixed, the optimal $\alpha_{k}$ can be expressed as
		\begin{eqnarray}
			\begin{aligned} \label{alpha}
				\alpha_{k}^{op}=\gamma_{k}.
			\end{aligned}
		\end{eqnarray}
	\end{mypro}
	\begin{IEEEproof}
		The partial derivative of $f_{2}(\pmb{\alpha},\!\mathbf{W},\!\pmb{\phi}_1\!,...,\!\pmb{\phi}_L)$ with respect to $\alpha_{k}$ can be written as
		\begin{eqnarray}
			\begin{aligned}\label{AP}
				\frac{\partial f_{2}(\pmb{\alpha},\mathbf{W},\pmb{\phi}_1,...,\pmb{\phi}_L)}{\partial \alpha_{k}}
				&=\frac{\gamma_{k}-\alpha_{k}}{(1+\gamma_{k})(1+\alpha_{k})}.
			\end{aligned}
		\end{eqnarray}
	
		By setting $\frac{\partial f_{2}(\pmb{\alpha},\mathbf{W},\pmb{\phi}_1,...,\pmb{\phi}_L)}{\partial \alpha_{k}}=0$, it is easy to find that $\alpha_{k}^{op}=\gamma_{k}$ maximizes $f_{2}(\pmb{\alpha},\mathbf{W},\pmb{\phi}_1,...,\pmb{\phi}_L)$, since $\partial f_{2}(\pmb{\alpha},\mathbf{W},\pmb{\phi}_1,...,\pmb{\phi}_L)/\partial \alpha_{k}>0$ when $\alpha_k<\gamma_{k}$ and $\partial f_{2}(\pmb{\alpha},\mathbf{W},\pmb{\phi}_1,...,\pmb{\phi}_L)/\partial \alpha_{k}<0$ when $\alpha_k>\gamma_{k}$.
	\end{IEEEproof}

	Besides, by defining $\pmb{\alpha}^{op}=[\alpha_1^{op},\alpha_2^{op},\cdots,\alpha_K^{op}]^T$ and substituting $\alpha_{k}^{op}=\gamma_{k}$ into \eqref{OBM1}, we have
	\begin{eqnarray} 
		\begin{aligned} \nonumber
			f_2(\pmb{\alpha}^{op},\mathbf{W},\pmb{\phi}_1,...,\pmb{\phi}_L)\!=\!\sum_{k=1}^{K} \log(1+\gamma_{k})\!=\!f(\mathbf{W},\pmb{\phi}_1,...,\pmb{\phi}_L),
		\end{aligned}
	\end{eqnarray}
 which guarantees the equivalence between the original problem $(P1)$ and the reformulated problem $(P2)$.

	\emph{2) The closed-form optimal solution for $\mathbf{W}$.}
	When $\pmb{\alpha}$ is fixed, the variable left for optimization in $(P2)$ is $\gamma_{k}$ only. Then the corresponding optimization problem can be rewritten as
	\begin{equation}
		\begin{split}\label{SINR}
			(P2.1)  \max_{\mathbf{W},\pmb{\phi}_1,...,\pmb{\phi}_L} &\quad\sum_{k=1}^{K} \frac{\tilde{\alpha}_{k} \gamma_{k}}{1+\gamma_{k}},\\
			\text{s.t.}  \quad &\quad|\phi_{l,n}|=1, ~l=1,2,...,L,~ n=1,2,...,N_l, \\
			& \quad\sum_{k=1}^{K}||\bm{w}_k||^2\leq P_{T},
\end{split}
\end{equation}
	where  $\tilde{\alpha}_{k}=1+\alpha_{k}$ for $k=1,2,...,K$.
	Substituting the expression (\ref{SINR_Ori}) of SINR into (\ref{SINR}), we can reformulate the objective function in $(P2.1)$ as
	\begin{eqnarray}
		\begin{aligned}
			\sum_{k=1}^{K} \frac{\tilde{\alpha}_{k} \gamma_{k}}{1+\gamma_{k}}= \sum_{k=1}^{K} \frac{\tilde{\alpha}_{k}|\mathbf{h}_{k}^H \bm{w}_{k}|^{2}}{\sum_{i=1}^{K}|\mathbf{h}_{k}^H\bm{w}_{i}|^{2}+\sigma_{0}^{2}}.
		\end{aligned}
	\end{eqnarray}
	Then, with given $\pmb{\alpha}$ and $\pmb{\phi}_1,...,\pmb{\phi}_L$, optimizing $\mathbf{W}$ becomes
	\begin{eqnarray}
		\begin{aligned}
			(P2.1a) \quad &\max_{\mathbf{W}} \quad f_{2.1a}(\mathbf{W})\!=\!\sum_{k=1}^{K}\frac{\tilde{\alpha}_{k}|\mathbf{h}_{k}^H \bm{w}_{k}|^2}{\sum_{i=1}^{K}|\mathbf{h}_{k}^H \bm{w}_{i}|^{2}\!+\!\sigma_{0}^{2}},\\
			&~\text{s.t.} \quad \quad\sum_{k=1}^{K}||\bm{w}_{k}||^{2}\leq P_{T}.
		\end{aligned} \label{P2.1}
	\end{eqnarray}
	
	 In \eqref{P2.1}, $(P2.1a)$ is a multiple-ratio fractional programming problem. Using the quadratic transform proposed in \cite{FP1,FP2}, $(P2.1a)$ can be reformulated as a biconvex optimization problem with a new objective function as
	\begin{eqnarray}
		\begin{aligned}\label{15}
			f_{2.1b}(\mathbf{W},\pmb{\xi})
			=&\sum_{k=1}^{K}\sqrt{\tilde{\alpha}_{k}} \left( \xi_{k}^{*}\mathbf{h}_{k}^H\bm{w}_{k}+\bm{w}_{k}^H\mathbf{h}_{k}\xi_{k}\right)\\
			&-\sum_{k=1}^{K}|\xi_{k}|^{2}\left(\sum_{i=1}^{K}|\mathbf{h}_{k}^H\bm{w}_{i}|^2+\sigma_{0}^{2}\right),
		\end{aligned}
	\end{eqnarray}
	where $\xi_{1},\xi_{2},...,\xi_{K}$ are the newly introduced complex auxiliary variables and $\pmb{\xi}=[\xi_{1},\xi_{2},...,\xi_{K}]^{T}$ is the auxiliary variable vector. Then, as proved in \cite{FP1} and \cite{FP2}, solving $(P3.1a)$ with respect to $\mathbf{W}$ is equivalent to solving the following  biconvex problem with respect to $\mathbf{W}$ and $\pmb{\xi}$, which can be further represented as  
	\begin{eqnarray}
		\begin{aligned}
			(P2.1b)\quad &\max_{\mathbf{W},\pmb{\xi}} \quad f_{2.1b}(\mathbf{W},\pmb{\xi}),\\
			& ~\text{s.t.}  \quad \sum_{k=1}^{K}||\bm{w}_{k}||^{2}\leq  P_{T}.
		\end{aligned}
	\end{eqnarray}
	By analyzing $(P2.1b)$, we have the following proposition regarding to its solution.
	\begin{mypro} 
		The optimal beamforming vector $\bm{w}_{k}$ can be updated by
		\begin{eqnarray}\label{ww}
			\begin{aligned}
				\bm{w}_{k}^{op}=\sqrt{\tilde{\alpha}_{k}}\xi_{k}^{op}\left(\sum_{i=1}^{K}|\xi_{i}^{op}|^{2}\mathbf{h}_{i}\mathbf{h}_{i}^H+\lambda^{op}\mathbf{I}_{M}\right)^{-1}\mathbf{h}_{k},
			\end{aligned}
		\end{eqnarray}
		where
		\begin{eqnarray} \label{beta}
			\begin{aligned}
				\xi_{k}^{op}=\frac{\sqrt{\tilde{\alpha}_{k}}\mathbf{h}_{k}^H\bm{w}_{k}}{\sum_{i=1}^{K}|\mathbf{h}_{k}^H\bm{w}_{i}|^{2}+\sigma_{0}^{2}}.
			\end{aligned}
		\end{eqnarray}
		In (\ref{ww}), $\lambda^{op}$ is the Lagrangian dual variable introduced for the power constraint $\sum_{k=1}^{K}||\bm{w}_{k}^{op}||^{2}=P_{T}$, which can be further expressed as
		\begin{eqnarray}
			\begin{aligned}\label{lamm}
				\lambda^{op}=\min\left\lbrace \lambda^{op} \geq 0:\sum_{k=1}^{K}||\bm{w}_{k}^{op}||^{2}=P_{T} \right\rbrace.
			\end{aligned}
		\end{eqnarray}
		
	\end{mypro}
	
	\begin{IEEEproof}
		See Appendix \ref{Appendix_A}.
	\end{IEEEproof}
 
 \emph{3) The closed-form optimal solution for phase shift vectors $\pmb{\phi}_1,\pmb{\phi}_2,...,\pmb{\phi}_L$.} For the optimized $\pmb{\alpha}$, the objective function in $(P2)$ is reduced to $(P2.1)$ in \eqref{SINR} as discussed above. By assuming that the beamforming matrix $\mathbf{W}$ at the BS side is fixed, we will optimize the reflection vector of each RIS iteratively, i.e, optimizing the reflection vector $\pmb{\phi}_l$, $l \in \lbrace 1,2,..,L\rbrace$, of RIS $l$ by assuming the rest are fixed. In the original expression of the equivalent channel for user $k$ in \eqref{Equ_channel}, we can see that the phase shift matrix of each RIS is coupled together due to the multi-hop transmission, which makes it impossible to solve them out. In order to overcome this, we perform channel reformulation as shown in the following, with which we can extract the phase shift matrix of each RIS as an individually objective variable. This will help us derive out the closed-form expression and thus they can be updated one by one with quite low complexity. Specifically, the reformulated channel model is represented as
 	\begin{equation} \label{Re_hk1}
 		\mathbf{h}_k^H=\bar{\pmb{\phi}_l}\mathbf{A}_l+\mathbf{b}_l.
 \end{equation} 
In \eqref{Re_hk1},  $\bar{\pmb{\phi}_l}$ is a row vector that is composed of $L-l+1$ reflecting vectors $\pmb{\phi}_l^H$ of RIS $l$, i.e., $ 
\bar{\pmb{\phi}_l}=[\pmb{\phi}_l^H,...,\pmb{\phi}_l^H]\in \mathbb{C}^{1 \times (L-l+1)N_l}$. The matrix $\mathbf{A}_l\in \mathbb{C}^{(L-l+1)N_l\times M}$ is represented as 
 	\begin{equation}
	\begin{split}	\label{A_l}
		\mathbf{A}_l=\left[\diag(\mathbf{g}_{l,k})\mathbf{R}_l,\diag(\mathbf{u}_{l+1})\mathbf{R}_l,
		...,\diag(\mathbf{u}_{L})\mathbf{R}_l
		\right]^T,
	\end{split}
\end{equation} 
where
\begin{equation} \label{R_l}
	\mathbf{R}_l=\left\{
	\begin{aligned}
		&\quad\mathbf{G}_1,\quad\quad\quad\quad\quad\quad\quad\quad  l=1 , \\
		&\!\!\prod_{i=l,l-1,...,2}\!\!\!\!\!\!\!\mathbf{H}_{i-1,i}\mathbf{\Phi}_{i-1}\mathbf{G}_1 , 
		\quad l=2,...,L,
	\end{aligned}
	\right.
\end{equation}
and 
\begin{equation} \label{u_i}
\mathbf{u}_i=\mathbf{g}_{i,k}\!\!\!\!\!\!\!\!\!\!\!\!\prod_{j=i,i-1,...,l+1}\!\!\!\!\!\!\!\!\!\!\!\!
\mathbf{\Phi}_{j}\mathbf{H}_{j-1,j},  \quad\quad i=l+1,l+2,...,L.
\end{equation}
In \eqref{Re_hk1}, $\mathbf{b}_l\in\mathbb{C}^{1\times M}$ is represented as
\begin{equation} \label{b_l}
	\mathbf{b}_l=\left\{
	\begin{aligned}
		\!\!\mathbf{0}, \quad\quad\quad\quad\quad\quad\quad\quad \quad\quad\quad\quad\quad\quad\quad\quad \quad &l=1 , \\
		\!\!\!\mathbf{g}_{1,k}\mathbf{\Phi}_1\mathbf{G}_1,\quad\quad\quad\quad\quad\quad\quad\quad\quad\quad\quad\quad\quad &l=2,\\
		\!\!\mathbf{g}_{1,k}\mathbf{\Phi}_1\mathbf{G}_1\!\!+\!\!\sum_{i=2}^{l-1}\mathbf{g}_{i,k}\!\!\!\!\!\!\!\!\!
		\prod_{j=i,i-1,...,2}\!\!\!\!\!\!\!\!\!\!\mathbf{\Phi}_j\mathbf{H}_{j-1,j}\mathbf{\Phi}_1\mathbf{G}_1,\quad &l=3,... ,L.
	\end{aligned}
	\right.
\end{equation}

 With the reformulated $\bm{h}_k^H$, we can rewrite the objective function in \eqref{SINR} with respect to $\pmb{\phi}_l$ as
\begin{equation}
\begin{split} \label{456}
f_{3.1a}(\pmb{\phi}_l)&=\sum_{k=1}^{K} \frac{\tilde{\alpha}_{k}|(\bar{\pmb{\phi}_{l}}\mathbf{A}_l+\bm{b}_l) \bm{w}_{k}|^{2}}{\sum_{i=1}^{K}|(\bar{\pmb{\phi}_{l}}\mathbf{A}_l+\bm{b}_l)\bm{w}_{i}|^{2}+\sigma_{0}^{2}}\\
&=\sum_{k=1}^{K} \frac{\tilde{\alpha}_k|\bar{\pmb{\phi}_l}\bm{z}_{l,k}+c_{l,k}|^2}
{\sum_{i=1}^K|\bar{\pmb{\phi}_l}\bm{z}_{l,i}+c_{l,i}|^2+\sigma_0^2}.
\end{split}
\end{equation}	
In \eqref{456}, the vector $\bm{z}_{l,i}\in \mathbb{C}^{(L-l+1)N_l \times 1}=\mathbf{A}_l\bm{w}_i$ and the complex number $c_{l,i}=\bm{b}_l\bm{w}_i$.  
	According to the quadratic transform proposed in \cite{FP1}, we further transform \eqref{456} into 
\begin{equation}
	\begin{split}\label{phi22}
			\hspace{-2mm}
			f_{3.2b}(\pmb{\phi}_l,\pmb{\varepsilon})\!\!=\!\!&\sum_{k=1}^{K}\sqrt{\tilde{\alpha}_k}
			\left(\varepsilon_k^*(\bar{\pmb{\phi}_l}\bm{z}_{l,k}+c_{l,k})\!+\!\!(\bm{z}_{l,k}^H\bar{\pmb{\phi}_l}^H\!+\!c_{l,k}^*)\varepsilon_k\right)\\
			&-\sum_{k=1}^{K}|\varepsilon_k|^2\left(\sum_{i=1}^{K}|\bar{\pmb{\phi}_l}\bm{z}_{l,i}+c_{l,i}|^2+\sigma_0^2\right),
\end{split}
\end{equation}
	where $\varepsilon_{1},\varepsilon_{2},...,\varepsilon_{K}$ are the newly introduced complex auxiliary variables and $\pmb{\varepsilon}=[\varepsilon_{1},\varepsilon_{2},...,\varepsilon_{K}]^{T}$ is the auxiliary variable vector. Then, as proved in \cite{FP1} and \cite{FP2}, optimizing $f_{3.2b}(\pmb{\phi}_l,\pmb{\varepsilon})$ with respect to $\pmb{\phi}_l$, $l=1,2,...L$, is equivalent to solving the following problem with respect to $\pmb{\phi}_l$ and $\pmb{\varepsilon}$, and we have the following objective function
	\begin{eqnarray}
		\begin{aligned}
			(P3.2b)\quad &\max_{\pmb{\phi}_l,\pmb{\varepsilon}} \quad 	f_{3.2b}(\pmb{\phi}_l,\pmb{\varepsilon}),\\
			& ~\text{s.t.}  \quad \quad|\phi_{l,n}|=1, \quad n=1,2,...,N_l.
		\end{aligned}
	\end{eqnarray}
	By analyzing $(P3.2b)$, we have the following conclusion regarding to the optimal solution for $\bm{\phi}_l$.
	\begin{mypro}
		The optimal expression $\phi_{l,n}, n=1,2,...,N_l$, of the reflecting vector $\pmb{\phi}_l$, $l=1,2,...,L$, is updated by
		\begin{eqnarray}
			\begin{aligned}\label{sphi}
				\phi_{l,n}^{op}=e^{-j\angle \eta_{n}},\quad n=1,2,...,N_l,
			\end{aligned}
		\end{eqnarray}
		where
		\begin{eqnarray}
		\begin{aligned}
		\eta_n=d_{l,n}-\sum_{j=1,j\neq n}^{(L-l+1)N_l}p_{n,j}\bar{\phi}^*_{l,j},
			\end{aligned}
		\end{eqnarray}
with $d_{l,n}$ being the $n$-th element of vector 
\begin{eqnarray}
\begin{aligned}
\bm{d}_{l}=
\sum_{k=1}^{K}\sqrt{\tilde{\alpha}_k}(\varepsilon_{k}^{op})^*\bm{z}_{l,k} 
- \sum_{k=1}^K|\varepsilon_k^{op}|^2\sum_{i=1}^{K}\bm{z}_{l,i}c_{l,i}^*.
	\end{aligned}
\end{eqnarray}
Among the denotations, $p_{n,j}$ represents the element at row $n$ and column $j$ of matrix
		$\mathbf{P}= \sum_{k=1}^K|\varepsilon_{k}^{op}|^2\sum_{i=1}^K\bm{z}_{l,i}\bm{z}_{l,i}^H$, and $\bar{\phi}_{l,j}$ represents the $j$-th element of the vector $\bar{\pmb{\phi}_l}$. Finally, $\varepsilon_{k}^{op}$ is proved to be
		\begin{eqnarray}
			\begin{aligned}\label{var}
				\varepsilon_{k}^{op}=\frac{\sqrt{\tilde{\alpha}_{k}}(\bar{\pmb{\phi}_l}\bm{z}_{l,k}+c_{l,k})}
				{\sum_{i=1}^{K}|\bar{\pmb{\phi}_l}\bm{z}_{l,i}+c_{l,i}|^{2}+\sigma_{0}^{2}}.
			\end{aligned}
		\end{eqnarray}
	\end{mypro}
	
	\begin{IEEEproof}
		See Appendix \ref{Appendix_B}.
	\end{IEEEproof} 

	\begin{algorithm}[t]
	\caption{:Cooperative beamforming design for multi-RIS-assisted multi-user systems.}
	\label{alg:SA}
	\begin{algorithmic}[1]\label{alg:algorithm2}
		\STATE \textbf{Initialization:} 
		Initialize $\pmb{\phi}_1,...,\pmb{\phi}_L$ that satisfy \eqref{PSC}. 
		Initialize $\mathbf{W}$ that satisfy \eqref{PC}. Set the threshold value $\ell$ and the maximum number of iterations $I_m$. Iteration time $i=1$.\\
		\STATE \textbf{Step 1:}  Update $\alpha_k$ according to (\ref{alpha}).
		\STATE \textbf{Step 2:}  Update $\bm{w}_k$, $\xi_{k}$ and $\lambda$ according to \eqref{ww}, \eqref{beta} and \eqref{lamm}, accordingly.
		\STATE \textbf{Step 3:}  Update $\phi_{l,n}$, $l=1,2,...L$, $n=1,2,...,N_l$, and $\varepsilon_k$ according to \eqref{sphi} and (\ref{var}), respectively.
		\STATE $i=i+1$.\\
		\textbf{UNTIL}\\ The objective function in \eqref{OBM1} satisfies $$f_{2}(\pmb{\alpha}^i,\mathbf{W}^i,\pmb{\phi}_1^i,...,\pmb{\phi}_L^i)-f_2(\pmb{\alpha}^{i-1},\mathbf{W}^{i-1},\pmb{\phi}_1^{i-1},...,\pmb{\phi}_L^{i-1})\leq \ell$$ or the iteration time $i=I_m$.
	\end{algorithmic}
\end{algorithm}
	
	Following the above approach, the phase shift of each RIS can be updated iteratively until the objective function converges.
	The overall algorithm to optimize the multi-RIS-assisted multi-user communication system is summarized in Algorithm 1, where we optimize the active beamforming matrix $\mathbf{W}$ and the phase shift vectors $\pmb{\phi}_1,..., \pmb{\phi}_L$ iteratively to maximize the sum rate. Algorithm 1 is guaranteed to converge (c.f. \textbf{Proposition 4}) and its computational complexity  mainly comes from the process of optimizing the active beamforming and updating the phase shifts. In the cooperative beamforming design processes,
	we first update $\alpha_k$, $k=1,2,...,K$, according to \eqref{alpha}, whose computational complexity is $\mathcal{O}(K)$. To update the active beamforming matrix $\mathbf{W}=[\bm{w}_{1},\bm{w}_{2},...,\bm{w}_{K}]$ at the BS side, we derive the closed-form optimal solution for $k$-th user's beamforming vector $\bm{w}_{k}$ as shown in \eqref{ww} with the required auxiliary variable $\xi_{k}^{op}$ shown in \eqref{beta}. The complexity for updating  $\xi_{k}^{op}$ is $\mathcal{O}(K)$ and the complexity for calculating $\mathbf{W}$ is $\mathcal{O}(M^3K)$, thereby the overall computational complexity for updating $\mathbf{W}$ is $\mathcal{O}(M^3K+K)$. Finally, for the phase shift design at each RIS, we iteratively update each RIS's phase shift vector $\pmb{\phi}_l$. The optimal
	expression $\phi_{l,n}, n=1,2,...,N_l$, of the reflecting vector $\pmb{\phi}_l$, $l=1,2,...,L$, is updated according to \eqref{sphi} with an auxiliary variable $\varepsilon_{k}^{op}$ calculated according to \eqref{var}. The complexity for updating $\varepsilon_{k}^{op}$, $k=1,2,...,K$, is $\mathcal{O}(K)$. And we iteratively update each RIS's phase shift vector  $\pmb{\phi}_l$ according \eqref{sphi} and its computational complexity is $\mathcal{O}(N_l)$. Thus, the aggregated complexity for getting $\pmb{\phi}_l$ is $\mathcal{O}\left(N_l+K\right)$. Since we deploy $L$ distributed RISs in the system, the overall complexity for updating RISs' phase shift is represented as $\mathcal{O}(KL+\sum_{l=1}^LN_l)$.
		Summarizing all above, the complexity of the proposed algorithm can be expressed as $\mathcal{O}\left(\left((M^3+L+2)K +\sum_{l=1}^L N_l\right)I_{1}\right)$, with $I_1$ denoting the required number of outer iterations.

	\begin{mypro}
		Algorithm 1 is guaranteed to converge.
	\end{mypro}
	\begin{IEEEproof}
		See Appendix \ref{Appendix_D}.
	\end{IEEEproof}
\section{Performance Evaluation}

In this section, performance evaluations are done to show the effectiveness of the proposed algorithm.
We first investigate the convergence property of the proposed cooperative beamforming algorithm, and then simulate its performance in terms of the complexity and the sum rate. For comparison purpose, we also adopt some existing cooperative beamforming algorithms for baselines. We investigate the impacts of the number of the reflecting elements and the channel estimation error on the sum rate. In our simulations,
we assume that there are
$L$ RISs (RIS 1, RIS 2,..., RIS $L$) deployed to assist the downlink communications, and RIS $l$ has $N_l$ reflecting elements. Since the RIS-reflected channel follows the product-distance path loss model, thereby we always deploy RIS 1 near the BS and RIS $L$ near the users to produce smaller multiplicative poss loss. Other RISs are deployed sequentially between RIS 1 and RIS $L$ to form multi-hop transmission links. Under the 3D Cartesian coordinate, the BS is deployed at $(0 \, \mathrm{m},0 \, \mathrm{m},0 \,\mathrm{m})$, the deployment location of RIS $l$, $l=1,2,...,L$, is represented by $(x_l \,\mathrm{m} ,y_l \,\mathrm{m},z_l\,\mathrm{m})$. The users are located within a target circular area $\mathcal{A}$, where $\mathcal{A}$ is centered at $A=(D \,\mathrm{m},0 \,\mathrm{m},0 \,\mathrm{m})$ with the radius of $r$ $\mathrm{m}$.
At the BS and user sides, we consider the widely used half-wavelength dipole antennas with corresponding maximum effective area $0.13 \lambda^2$ \cite{antenna}, and the size of each RIS element is $\lambda/5 \times \lambda/5$ \cite{RIS_size}. By considering the physical area of the RIS element as the effective area, we can calculate the corresponding free-space path loss of each channel according to Friis transmission formula \cite{Friis}.
 Other system parameters are summarized in Table \ref{table_par} according to \cite{channel_model}. All presented results are obtained by averaging $1000$ independent realizations of the users' locations and the wireless channels.
\renewcommand\arraystretch{1.3}
\vspace{-0.5cm}
\begin{table}[tb]
	\centering
	\caption{\textsc{Simulation Setups}}
		\label{table_par}
		\begin{tabular}{p{5cm}<{\centering}|p{2cm}<{\centering}}
			\hline
		The central location $A$ of the user cluster  &  $(10\, \mathrm{m}, 0\, \mathrm{m}, 0\, \mathrm{m})$  \\
			\hline 
		The radius $r$ of the target area  & $8$ $\mathrm{m}$\\
			\hline
		The	transmit power $P_T$ at the BS  &10 dBW	\\
			\hline	
		The Rician factor for all the channels &  3\\
			\hline
			The noise power $\sigma_0^2$ &-110 dBm\\
			\hline
			The carrier frequency & 2.4 GHz\\
		\hline
		\end{tabular}	
\end{table}

\subsection{The Convergence Property of the Proposed Cooperative Beamforming Algorithm}
\begin{figure}[t]
	\centering
	\vspace{-0.3cm}
	\includegraphics[width=0.4\textwidth]{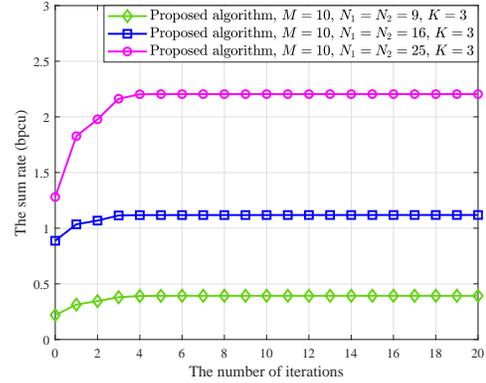}\\
	\caption{The convergence property of the proposed beamforming algorithm. There are $L=2$ RISs (RIS 1 and RIS 2). RIS 1 is deployed at $(1\, \mathrm{m}, 0\, \mathrm{m}, 3\, \mathrm{m})$ and RIS 2 is deployed at $(9\, \mathrm{m}, 0\, \mathrm{m}, 3\, \mathrm{m})$.  }
	\label{fig:C2}
\end{figure}
\begin{figure}[t]
	\centering
		\vspace{-0.4cm}
	\includegraphics[width=0.4\textwidth]{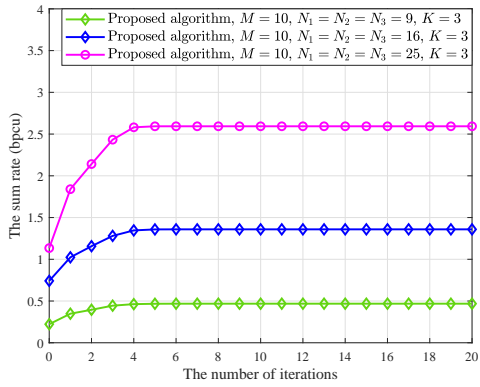}\\
	\caption{The convergence property of the proposed beamforming algorithm. There are $L=3$ RISs (RIS 1, RIS 2 and RIS 3). RIS 1 is deployed at $(1\, \mathrm{m}, 0\, \mathrm{m}, 3\, \mathrm{m})$, RIS 2 is deployed at $(5\, \mathrm{m}, 0\, \mathrm{m}, 3\, \mathrm{m})$ and RIS 3 is deployed at $(9\, \mathrm{m},0\, \mathrm{m},3\, \mathrm{m})$. }
	\label{fig:C22}
\end{figure}

The convergence property of the proposed cooperative beamforming algorithm is first investigated under various system settings. The results are illustrated in Figs. \ref{fig:C2} and \ref{fig:C22}. Specifically, Fig. \ref{fig:C2} shows the convergence property of the proposed beamforming algorithm for $L=2$ RISs (RIS 1 and RIS 2), where RIS 1 is deployed at $(1\, \mathrm{m}, 0\, \mathrm{m}, 3\, \mathrm{m})$ and RIS 2 is deployed at $(9\, \mathrm{m}, 0\, \mathrm{m}, 3\, \mathrm{m})$. In the simulations, the numbers of the reflecting elements at these two RISs are set to be equal, i.e., $N_1=N_2$. It can be observed in Fig. \ref{fig:C2} that the proposed beamforming algorithm converges within several iterations in cases with $N_1=N_2=9$, $N_1=N_2=16$ and $N_1=N_2=25$, respectively. This verifies \textbf{Proposition 4} and the fast convergence implies a relatively low computational complexity of the proposed cooperative beamforming algorithm.

Fig. \ref{fig:C22} shows the convergence behavior of the proposed beamforming algorithm for $L=3$ RISs, where RIS 1 is deployed at $(1\, \mathrm{m}, 0\, \mathrm{m}, 3\, \mathrm{m})$, RIS 2 is deployed at $(5\, \mathrm{m}, 0\, \mathrm{m}, 3\, \mathrm{m})$ and RIS 3 is deployed at $(9\, \mathrm{m}, 0\, \mathrm{m}, 3\, \mathrm{m})$. In the simulations, the numbers of the reflecting elements at these three RISs are also set to be the same, i.e., $N_1=N_2=N_3$. Similarly, for the 3 RISs case, the proposed algorithm also converges within several  iterations. Comparing with Fig. \ref{fig:C2}, we observe that although the number of the deployed RISs increases,
our proposed algorithm can still converge quickly. This implies that the proposed beamforming algorithm does have a good expandability.
\vspace{-0.3cm}
\subsection{Comparisons with the Existing Cooperative Beamforming Algorithms}
For comparison, we choose the semidefinite relaxation (SDR)-based and alternating direction method of multipliers (ADMM)-based algorithms as benchmarks. We choose these two algorithms because they are two different methods handling non-convex rank-one constraint optimization problems, which is also the constraint in our design. The SDR-based method can directly relax the rank-one constraint, while ADMM-based algorithm transfers the non-convex constraint to the objective function by adding a penalty factor. Specifically,	

\begin{itemize}
	\item
		The SDR-based algorithm \cite{N2} can relax the non-convex rank-one
		constraint of the RIS's phase shifts, and transform the original objective function into a convex semidefinite program (SDP) to make such problems easily be solved through existing convex optimization methods.
	
	\item 
		The ADMM-based algorithm \cite{wang2019global} solves such problem by introducing an auxiliary vector $\bm{q}_l$, then adding a penalty term into the objective function. By doing so, the objective function can be reformulated into Lagrange dual problem and solved by gradient descent method.
		       
\end{itemize}

\begin{figure}[tp]
	\centering
	\vspace{-0.3cm}
	\includegraphics[width=0.4\textwidth]{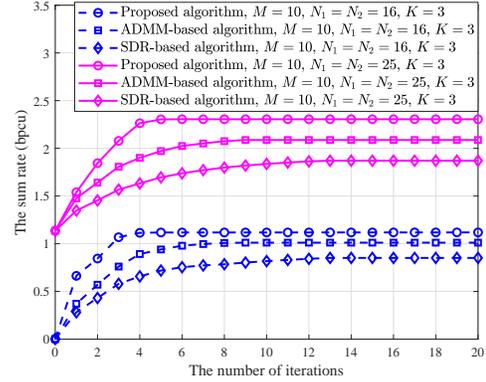}\\
	\caption{The convergence properties of different algorithms.There are $L=2$ RISs (RIS 1 and RIS 2). RIS 1 is deployed at $(1\, \mathrm{m}, 0\, \mathrm{m}, 3\, \mathrm{m})$ and RIS 2 is deployed at $(9 \, \mathrm{m}, 0\, \mathrm{m}, 3\, \mathrm{m})$. $M=10$ and $K=3$.}
	\label{fig:COM_Convergence}
\end{figure}

The computational complexity of these algorithms are listed in Table \ref{table_complexity}. We can observe from Table \ref{table_complexity} that the complexity of the proposed algorithm linearly increases with the number of the RIS's reflecting elements, which indicates that the proposed algorithm can be easily extended to systems with more RISs. On the contrary, the SDR- and ADMM-based algorithms cannot be directly applied to systems with large-scale RISs because of the prohibitively high computational complexity.
For comparsion purpose, we show the convergence property of these algorithms in Fig. \ref{fig:COM_Convergence}.
It is observed that the convergence speed of the proposed algorithm is obviously faster than the SDR- and ADMM-based algorithms. 
The reason is that in each iteration, our proposed algorithm is updated with the closed-form optimal solutions, which not only improves the sum rate, but also reduces the computational complexity. 
\begin{table}[t]  
	\newcommand{\tabincell}[2]{\begin{tabular}{@{}#1@{}}#2\end{tabular}}
	\centering
	\caption{\textsc{Computational Complexity for Different Cooperative Beamforming Algorithms}} \label{table_complexity}
	\begin{threeparttable}
		\begin{tabular}{m{2.85cm}<{\centering}|m{5.2cm}<{\centering}}
			
			\hline
			\textbf{Algorithm} & \textbf{Computational complexity for $L=2$ RISs} \\
			
			\hline
			
			Proposed algorithm 	&  $\mathcal{O}\left((M^3K+N_1+N_2+4K)I_{1}\right)$\\
			
			\hline
			
			ADMM-based algorithm  & $\mathcal{O}\left((M^3K+N_1^3+N_2^3+2K)I_2\right)$\\
			
			\hline
			
			SDR-based algorithm   &$\mathcal{O}\left((M^3K+N_1^6+N_2^6+2K)I_3\right)$\\
			\hline
		\end{tabular}			
		\begin{tablenotes}
			\footnotesize
			\item[*]  $I_n$ with $n=1,2,3$ is the number of outer iterations when the algorithm converges.
		\end{tablenotes}	
	\end{threeparttable}
\end{table}
\begin{figure}[tp]
	\centering
	\vspace{-0.6cm}
	\includegraphics[width=0.4\textwidth]{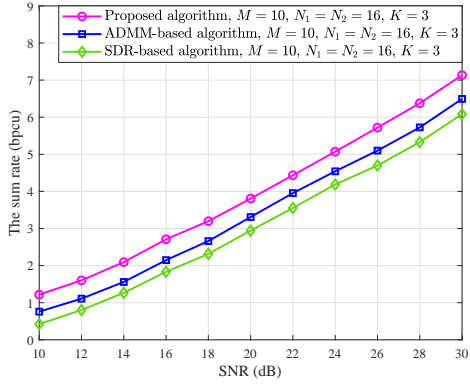}\\
	\caption{The sum rate comparison for different SNR values. There are $L=2$ RISs (RIS 1 and RIS 2). \\RIS 1 is deployed at $(1\, \mathrm{m}, 0\, \mathrm{m}, 3\, \mathrm{m})$ and RIS 2 is deployed at $(9 \, \mathrm{m}, 0\, \mathrm{m}, 3\, \mathrm{m})$. $M=10$ and $K=3$.}
	\label{fig:C61}
\end{figure}

Fig. \ref{fig:C61} shows the sum rate performance of these cooperative beamforming algorithms for different SNR values. It is observed that as the SNR increases, the proposed algorithm can always achieve the highest sum rate. This is owing to that the closed-form optimal solutions of the individual variables are obtained in each iteration. Based on the above observations in Table \ref{table_complexity}, Fig. \ref{fig:COM_Convergence} and Fig. \ref{fig:C61}, we claim that the proposed algorithm can provide higher sum rate with much lower computational complexity.

\begin{figure}[tp]
	\centering
	\vspace{-0.3cm}
	\includegraphics[width=0.4\textwidth]{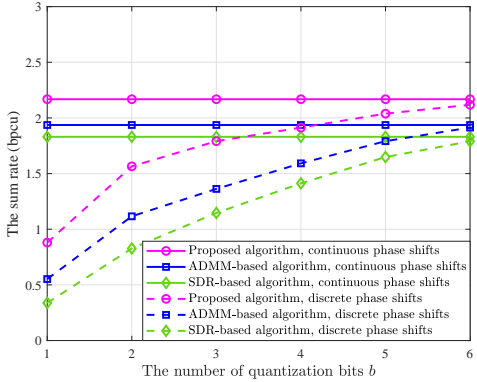}\\
	\caption{The sum rate performance with different numbers of quantization bits $b$.
		 There are $L=2$ RISs (RIS 1 and RIS 2). \\RIS 1 is deployed at $(1\, \mathrm{m}, 0\, \mathrm{m}, 3\, \mathrm{m})$ and RIS 2 is deployed at $(9 \, \mathrm{m}, 0\, \mathrm{m}, 3\, \mathrm{m})$, $N_1=N_2=25$, $M=10$ and $K=3$.}
	\label{fig:discrete_phase}
\end{figure}

Considering the deployment cost of RIS, is it possible to consider discrete values for the RIS' phase shifts in the practical implementations. In discrete phase case, we refer to $b$ as the number of quantization bits and $B=2^b$ as the discrete phase shift level. Thus the set of discrete phase-shift values at each element is given by
$
\mathcal{F}=\lbrace
0,\Delta\theta,...,(B-1)\Delta\theta
\rbrace,
$
where  $\Delta\theta=2\pi/B$. To investigate the impact of the discrete phase shift level on the sum rate performance, we depicts the sum rate of all users versus the number of quantization bits $b$ in Fig. \ref{fig:discrete_phase}. As the number of quantization bits increases, the sum rate obtained with discrete phase shifts approaches that in the continuous case and our proposed algorithm can always achieve the best sum rate performance compared with the benchmark schemes.

\subsection{Impacts of the Number of Reflecting Elements and the Channel Estimation Errors}
\begin{figure}[tp]
	\centering
	\vspace{-0.6cm}
	\includegraphics[width=0.4\textwidth]{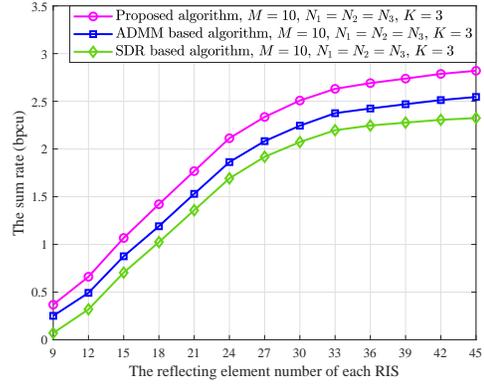}\\
	\caption{The sum rate with different reflecting element numbers for $L=3$ RISs (RIS 1, RIS 2 and RIS 3).\\ RIS 1 is deployed at $(1\, \mathrm{m}, 0\, \mathrm{m}, 3\, \mathrm{m})$, RIS 2 is deployed at $(5 \, \mathrm{m}, 0\, \mathrm{m}, 3\, \mathrm{m})$ and RIS 3 is deployed at $(9\, \mathrm{m},0\, \mathrm{m},3\, \mathrm{m})$.}
	\label{fig:C6}
\end{figure}
The performance of the RIS-assisted communication systems greatly depends on the number of the reflecting elements. To investigate the dependency between them, we simulate the sum rate performance of the proposed algorithm for different numbers of the RISs' reflecting elements. In the simulations, $L=3$ RISs (RIS 1, RIS 2 and RIS 3) are deployed, and these three RISs have the same number of reflecting elements. The SDR- and ADMM-based cooperative beamforming  algorithms are also simulated for comparison.
For $M = 10$, $K = 3$, the simulation results are shown in Fig. \ref{fig:C6}.
It is observed that as the number of the RISs' reflecting elements increases,
 the sum rates of all the three algorithms increase and the proposed algorithm always achieves the highest sum rate compared with the chosen baseline algorithms. 
We can also observe from Fig. \ref{fig:C6} that
the improvement of the sum rate is limited by the inter-user interference. Since the RISs do not have any signal processing ability and they reflect the overall aggregated signals, thereby as the reflecting element number continuously increases, the sum rate performance tends to converge to ceiling.

\begin{figure}[tp]
	\centering
		\vspace{-0.2cm}
	\includegraphics[width=0.4\textwidth]{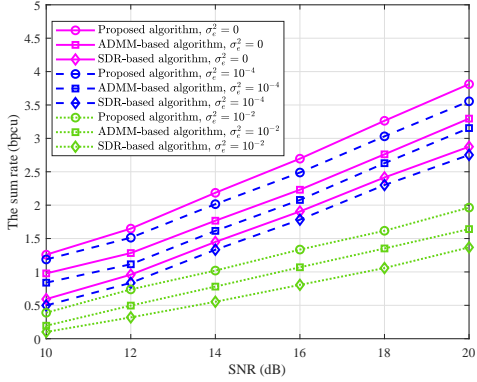}\\
	\caption{The sum rate comparison for different channel estimation errors. There are $L=2$ RISs (RIS 1 and RIS 2).\\ RIS 1 is deployed at $(1\, \mathrm{m}, 0\, \mathrm{m}, 3\, \mathrm{m})$ and RIS 2 is deployed at $(9 \, \mathrm{m}, 0\, \mathrm{m}, 3\, \mathrm{m})$. $N_1=N_2=16$, $M=10$ and $K=3$.}
	\label{fig:C62}
\end{figure}

For the beamforming algorithms, imperfect CSI will affect the final sum rate performance. To show how the channel estimation ambiguity impacts the system performance when applying our proposed algorithm, we conduct simulations to evaluate the sum rate under different channel estimation errors. 
Specifically, we model all the estimated channel vectors or matrices, i.e., BS-RIS 1 channel, inter-RIS channels and RIS-user channels, as $\hat{\bm{r}}=\bm{r}+\bm{r}_e$ or $\hat{\bm{\Upsilon}}=\bm{\Upsilon}+\bm{\Upsilon}_e$ with $\bm{r}$ and $\bm{\Upsilon}$ representing the actual channel, $\bm{r}_e$ and $\bm{\Upsilon}_e$ denoting the channel estimation error.  Each element in $\bm{r}_e$ and $\bm{\Upsilon}_e$ is chosen from $\mathcal{CN}(0,\sigma_e^{2})$. And $\sigma_e^{2}$ can reflect the level of the channel estimation ambiguity. when $\sigma_e^{2}$ is bigger, it indicates that the channel estimation error is larger, and vice versa indicates that the channel estimation error is small.
It is observed from Fig. \ref{fig:C62} that as the channel estimation error increases, the sum rate obtained by all the three algorithms decrease and our proposed algorithm always achieves the best sum rate. It is also shown that when the channel estimation error is small, e.g., $\sigma_e^{2}=10^{-4}$, the sum rate achieved by our proposed algorithm is even better than that of the ADMM- and SDR-based algorithms with perfect CSI.
\subsection{Comparisons with Existing Design Strategies for Equal Total Reflecting Elements}
\begin{figure}[tp]
	\centering
	\vspace{-0.3cm}
	\includegraphics[width=0.4\textwidth]{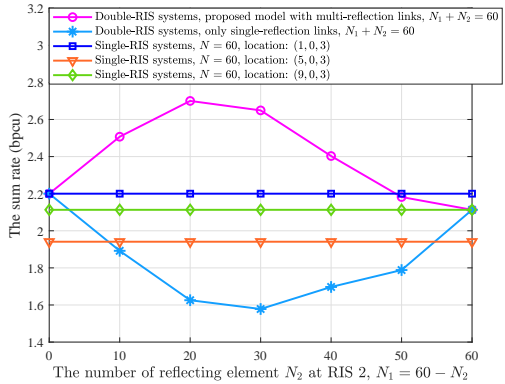}\\
	\caption{The sum rate comparison for the proposed channel model and existing baseline schemes with equal total reflecting elements. In double-RIS systems, RIS 1 is deployed at $(1\, \mathrm{m},0\, \mathrm{m},3\, \mathrm{m})$, RIS 2 is deployed at $(9\, \mathrm{m},0\, \mathrm{m},3\, \mathrm{m})$. In single-RIS systems, the RIS is deploying at three different locations, i.e., $(1\, \mathrm{m},0\, \mathrm{m},3\, \mathrm{m})$, $(5\, \mathrm{m},0\, \mathrm{m},3\, \mathrm{m})$ and $(9\, \mathrm{m},0\, \mathrm{m},3\, \mathrm{m})$.}
	\label{fig:C7}
\end{figure}

To show how much sum rate gains can be obtained by the proposed transmission strategy, we also compare the sum rate for double- and single-RIS-assisted systems with equal total reflecting elements, the results are shown in Fig. \ref{fig:C7}. Specifically, for the double-RIS systems, we compare the proposed multi-reflection transmission model with the baseline scheme that only studies the single reflection links, i.e., only BS-RIS 1-user links and BS-RIS 2-user links.
And in these two double-RIS scenarios, we assume RIS 1 has $N_1$ reflecting elements and deployed at $(1\, \mathrm{m},0\, \mathrm{m},3\, \mathrm{m})$, RIS 2 has $N_2$ reflecting elements and deployed at $(9\, \mathrm{m},0\, \mathrm{m},3\, \mathrm{m})$. 
We also compare the proposed cooperative transmission strategy with single-RIS scheme for equal total reflecting elements.
In single-RIS scenario, the RIS has $N=N_1+N_2$ reflecting elements and for thorough comparison, the performance is simulated for the $N$ elements in different locations, $(1\, \mathrm{m},0\, \mathrm{m},3\, \mathrm{m})$, $(5\, \mathrm{m},0\, \mathrm{m},3\, \mathrm{m})$ and $(9\, \mathrm{m},0\, \mathrm{m},3\, \mathrm{m})$. 
We can observe that it is beneficial to adopt the proposed cooperative transmission scheme in most cases, since the multi-reflection link established through RIS 1 and RIS 2 can bring significant multiplicative beamforming gains compared with other benchmark schemes.
We also observe that the number of the reflecting elements at RIS 1 and RIS 2 have great effect on the sum rate, and the peak sum rate is obtained when $N_1=40$ and $N_2=20$ for the simulated scenario.
In particular, for the baseline scheme that only considers the single-reflection links of these two distributed RISs, due to the smaller number of reflecting elements at each RIS and the lack of multiplicative beamforming gains, its performance is even worse than single-RIS schemes.  

\begin{figure}[tp]
	\centering
	\vspace{-0.2cm}
	\includegraphics[width=0.4\textwidth]{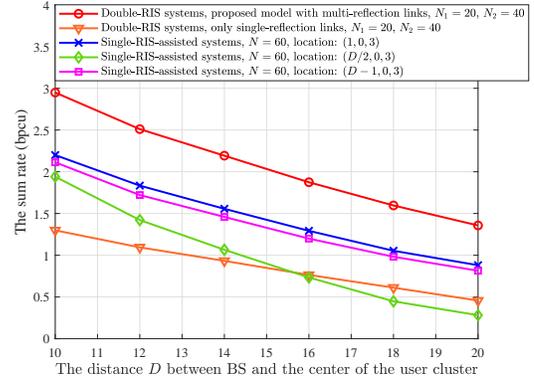}\\
	\caption{The sum rate comparison for the proposed channel model and existing baseline schemes when $D$ increasing. In double-RIS scenario, RIS 1 is deployed at $(1\, \mathrm{m},0\, \mathrm{m},3\, \mathrm{m})$, RIS 2 is deployed at $(D-1 \, \mathrm{m},0\, \mathrm{m},3\, \mathrm{m})$. In single-RIS scenario, the RIS is deployed at three different locations, i.e., $(1\, \mathrm{m},0\, \mathrm{m},3\, \mathrm{m})$, $(D/2\, \mathrm{m},0\, \mathrm{m},3\, \mathrm{m})$ and $(D-1\, \mathrm{m},0\, \mathrm{m},3\, \mathrm{m})$.}
	\label{fig:C8}
\end{figure}

To further verify the superiority of the proposed multi-RIS cooperative transmission scheme, we study the sum rate performance when the distance $D$ between the center of the user cluster and the BS varies.
In Fig. \ref{fig:C8}, we show the sum rate achieved by double- and single-RIS-assisted systems.
In double-RIS scenarios, we also compare the proposed multi-reflection transmission model with the baseline scheme that only studies the single reflection links, i.e., only BS-RIS 1-user links and BS-RIS 2-user links.
And for these two schemes, we use the same systems setup, i.e.,
RIS 1 has $N_1$ reflecting elements and is deployed at $(1\, \mathrm{m},0\, \mathrm{m},3\, \mathrm{m})$, RIS 2 has $N_2$ reflecting elements and is deployed at $(D-1 \, \mathrm{m},0\, \mathrm{m},3\, \mathrm{m})$. 
For the comparative single-RIS case, we consider equal total reflecting elements, the RIS has $N=N_1+N_2$ reflecting elements, and again the system is simulated when the RIS is depoyed at different locations, i.e., $(1\, \mathrm{m},0\, \mathrm{m},3\, \mathrm{m})$, $(D/2\, \mathrm{m},0\, \mathrm{m},3\, \mathrm{m})$ and $(D-1\, \mathrm{m},0\, \mathrm{m},3\, \mathrm{m})$.
It can be observed from Fig. \ref{fig:C8} that as $D$ increases, the sum rate of both the single- and double-RIS assisted systems decrease due to the increasing transmit distance. However, the proposed cooperative multi-RIS transmission scheme can always achieve higher sum rate with the help of the multiplicative beamforming gains provided by the multi-reflection links. 
\subsection{The Impact of the Number of Distributed RISs and the Deployment Location on Sum Rate}
\begin{figure}[tp]
	\centering
	\vspace{-0.2cm}
	\includegraphics[width=0.4\textwidth]{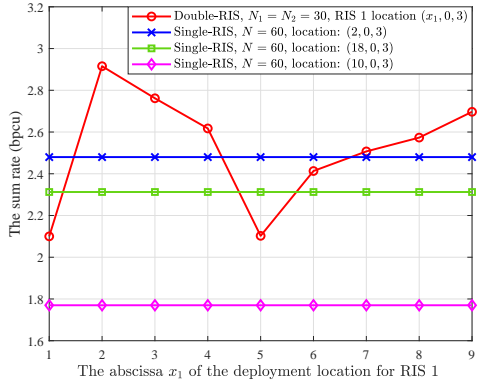}\\
	\caption{The sum rate performance under different RISs' deployment locations. In double-RIS scenario, RIS 1 is deployed at $(x_1 \, \mathrm{m}, 0 \,\mathrm{m}, 3\, \mathrm{m})$ while RIS 2 is deployed at $\left( (20-x_1)\, \mathrm{m},0\, \mathrm{m}, 3\,\mathrm{m}\right)$. In single-RIS scenario, the RIS is deployed at $(2 \,\mathrm{m}, 0 \,\mathrm{m}, 3 \,\mathrm{m})$, $(10 \,\mathrm{m}, 0\, \mathrm{m}, 3\, \mathrm{m})$ and $(18\, \mathrm{m}, 0\, \mathrm{m}, 3\,\mathrm{m})$, respectively.}
		\label{fig:deployment_location}
\end{figure}
To gain some useful insights for the optimal deployment locations of the RISs, we simulate the sum rate performance under different RISs' deployment locations in Fig. \ref{fig:deployment_location} with $L=2$ distributed RISs. In doing so, we consider a double-RIS setup. At the beginning, RIS $1$ and RIS $2$ are closed to the base station and the user cluster, respectively, then they gradually move closer to each other. The results in Fig. \ref{fig:deployment_location} show that the placement of the RISs have great influences on the sum rate performance. For the double-RIS scenario, the best placement for RIS 1 and RIS 2 are the points of $(2 \,\mathrm{m}, 0\, \mathrm{m}, 3\, \mathrm{m})$ and $(18\, \mathrm{m}, 0\, \mathrm{m}, 3\, \mathrm{m})$, since the RIS-reflected channel follows the product-distance path loss model, it is better to make some channels have longer transmission distance while some channels have shorter transmission distance.
	As RIS 1 and RIS 2 are getting closer to each other, the sum rate performance first decreases then gradually increases. This is because when the two RISs getting closer, the transmission distance between the BS and RIS 1 becomes longer, resulting in heavier channel pass loss, which severely restricts the sum rate performance. However, when RIS 1 and RIS 2 are getting even closer, the inter-path between these two RISs can provide considerable multiplicative beamforming gain, thereby improves the sum rate. For single-RIS case, when the RIS is deployed close to the base station, the sum rate performance will be better, but when the RIS is deployed in the middle between the BS and the user cluster, the sum rate will be worse due to the heavy product-distance path loss.
\begin{figure}[tp]
	\centering
	\vspace{-0.2cm}
	\includegraphics[width=0.4\textwidth]{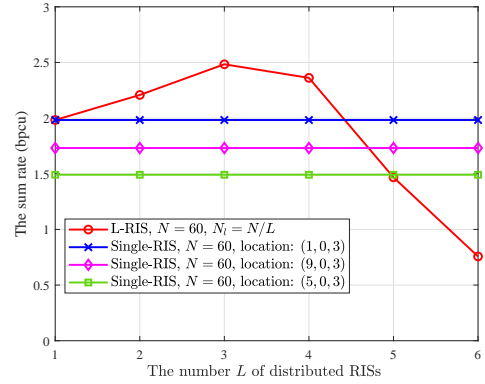}\\
	\caption{The sum rate performance under different distributed-RIS number $L$ with fixed total available reflecting elements $N=60$, each RIS has $N_l=N/L$ reflecting elements. }
	\label{fig:RIS_number}
\end{figure}

To investigate the impact of the distributed-RIS number on the sum rate performance, we simulated the sum rate performance under different distributed-RIS numbers in Fig. \ref{fig:RIS_number}. 
Specifically, with fixed total available reflecting elements $N=60$,  we change the number $L$ of the distributed RISs, and each RIS has $N_l=N/L$ reflecting elements.  
For L-RIS scenario, when $L=1$, the RIS is deployed at $(1\, \mathrm{m},0\, \mathrm{m},3\, \mathrm{m})$; when $L=2$, these two RISs are deployed at $(1\, \mathrm{m},0\, \mathrm{m},3\, \mathrm{m})$ and $(9\, \mathrm{m},0\, \mathrm{m},3\, \mathrm{m})$, respectively, and each RIS has $N_1=N_2=30$ reflecting elements. When $L=3$, RIS 1, RIS 2 and RIS 3 are deployed at $(1\, \mathrm{m},0\, \mathrm{m},3\, \mathrm{m})$, $(5\, \mathrm{m},0\, \mathrm{m},3\, \mathrm{m})$ and $(9\, \mathrm{m},0\, \mathrm{m},3\, \mathrm{m})$, each RIS has $N_1=N_2=N_3=20$ reflecting elements. When $L=4$, these four RISs are deployed at $(1\, \mathrm{m},0\, \mathrm{m},3\, \mathrm{m})$, $(3\, \mathrm{m},0\, \mathrm{m},3\, \mathrm{m})$, $(6\, \mathrm{m},0\, \mathrm{m},3\, \mathrm{m})$ and $(9\, \mathrm{m},0\, \mathrm{m},3\, \mathrm{m})$, each RIS has $N_1=N_2=N_3=N_4=15$ reflecting elements. When $L=5$, these RISs are deployed at $(1\, \mathrm{m},0\, \mathrm{m},3\, \mathrm{m})$, $(3\, \mathrm{m},0\, \mathrm{m},3\, \mathrm{m})$, $(5\, \mathrm{m},0\, \mathrm{m},3\, \mathrm{m})$, $(7\, \mathrm{m},0\, \mathrm{m},3\, \mathrm{m})$ and $(9\, \mathrm{m},0\, \mathrm{m},3\, \mathrm{m})$, each RIS has $N_1=N_2=N_3=N_4=N_5=12$ reflecting elements. When $L=6$, these RISs are deployed at $(0\, \mathrm{m},0\, \mathrm{m},3\, \mathrm{m})$, $(2\, \mathrm{m},0\, \mathrm{m},3\, \mathrm{m})$, $(4\, \mathrm{m},0\, \mathrm{m},3\, \mathrm{m})$, $(6\, \mathrm{m},0\, \mathrm{m},3\, \mathrm{m})$, $(8\, \mathrm{m},0\, \mathrm{m},3\, \mathrm{m})$ and $(10\, \mathrm{m},0\, \mathrm{m},3\, \mathrm{m})$, each RIS has $N_1=N_2=N_3=N_4=N_5=N_6=10$ reflecting elements. From Fig. \ref{fig:RIS_number}, we observe that when the total number of the available reflecting elements is fixed,
the number of distributed RISs has a significant impact on sum rate performance. When $L$ increasing, the sum rate first increases due to the multiplicative beamforming gains obtained through multi-reflection channels, and then decreases. This is because with fixed total available reflecting elements, the continuous growth of $L$ will lead to a decrease in the number of available reflecting elements per RIS, resulting in a smaller passive beamforming gain at each RIS and finally a decrease of the sum rate performance.

\section{Conclusions}
In this paper, we have investigated 
the cooperative beamforming design for multi-RIS-assisted multi-user communication systems. We have formulated a downlink sum rate optimization problem for arbitrary reflection links. 
The formulated problem is diffcult to solve due to the non-convexity and the interactions among the decision variables, we have decoupled the original problem into three disjoint subproblems. By introducing appropriate auxiliary variables in each subproblem, we have derived the closed-form expressions for the optimization decision variables. Based on these closed-form expressions, we have designed a low-complexity cooperative beamformig algorithm. We have carried out extensive simulation studies and shown that the proposed algorithm has linear complexity with the number of RISs' reflecting elements. We have also verified that the proposed algorithm achieves higher sum rate with lower complexity
compared with other commonly used alternative approaches. The results have also unveiled that appropriately distributing the reflecting elements among multiple RISs performs better than deploying them at one single RIS in terms of achievable  sum rate. 
	
	\appendices
	\section{Proof of Proposition 2}
	\label{Appendix_A}
	The optimization problem is
		\begin{eqnarray}
	\begin{aligned}
	(P2.1b)\quad &\max_{\mathbf{W},\pmb{\xi}} \quad f_{2.1b}(\mathbf{W},\pmb{\xi}),\\
	& ~\text{s.t.}  \quad \sum_{k=1}^{K}||\bm{w}_{k}||^{2}\leq  P_{T}.
	\end{aligned}
	\end{eqnarray}
	with 
		\begin{eqnarray}
	\begin{aligned}
	f_{2.1b}(\mathbf{W},\pmb{\xi})
	=&\sum_{k=1}^{K}\sqrt{\tilde{\alpha}_{k}} \left( \xi_{k}^{*}\mathbf{h}_{k}^H\bm{w}_{k}+\bm{w}_{k}^H\mathbf{h}_{k}\xi_{k}\right)\\
	&-\sum_{k=1}^{K}|\xi_{k}|^{2}\left(\sum_{i=1}^{K}|\mathbf{h}_{k}^H\bm{w}_{i}|^2+\sigma_{0}^{2}\right).
	\end{aligned}
	\end{eqnarray}
	 
	First, when $\mathbf{W}$ is fixed, the partial gradient of $f_{2.1b}(\mathbf{W},\pmb{\xi})$ with respect to $\xi_{k}$ can be expressed as
	\begin{eqnarray}
		\begin{aligned}
			\frac{\partial f_{2.1b}(\mathbf{W},\pmb{\xi})} {\partial \xi_{k}}\!=\!2\sqrt{\tilde{\alpha}_{k}}\mathbf{h}_{k}^H\mathbf{w}_{k}\!-\!2\xi_{k}\sum_{i=1}^{K}|\mathbf{h}_{k}^H\bm{w}_{i}|^{2}\!-\!2\xi_{k}\sigma_{0}^{2}.
		\end{aligned}
	\end{eqnarray}
	By solving $\partial f_{2.1b}(\mathbf{W},\bm{\beta})/ \partial \xi_{k}=0$, the optimal $\xi_{k}^{op}$ can be derived as shown in \eqref{beta}. Next,
	with fixed $\xi_{k}^{op}$, $f_{2.1b}(\mathbf{W},\pmb{\xi})$ can be rewritten as
	\begin{eqnarray}
		\begin{aligned}\label{20}
			f_{2.1b}(\mathbf{W},\pmb{\xi}^{op})=&\sum_{k=1}^{K}\sqrt{\tilde{\alpha}_{k}}\left(\xi_{k}^{op^{*}}\mathbf{h}_{k}^H\bm{w}_{k}+\bm{w}_{k}^{H}\mathbf{h}_{k}\xi_{k}^{op}\right)\\
			&-\sum_{k=1}^{K}|\xi_{k}^{op}|^{2} \sum_{i=1}^{K}\bm{w}_{i}^{H}\mathbf{h}_{k}\mathbf{h}_{k}^H\bm{w}_{i}.
		\end{aligned}
	\end{eqnarray}
	Considering the power constraint $\sum_{k=1}^{K}||\bm{w}_{k}||^{2}= P_{T}$, we write the Lagrangian function of $f_{2.1b}(\mathbf{W},\pmb{\xi}^{op})$ as
	\begin{eqnarray}
		\begin{aligned}
			&f_{2.1b}^{(L)}(\lambda, \mathbf{W}, \pmb{\xi}^{op})\!\!=\!\!\sum_{k=1}^{K}\sqrt{\tilde{\alpha}_{k}}\left(\xi_{k}^{op^{*}}\mathbf{h}_{k}^H\bm{w}_{k}\!+\!\bm{w}_{k}^{H}\mathbf{h}_{k}\xi_{k}^{op}\right)\\
			&-\sum_{k=1}^{K}|\xi_{k}^{op}|^{2} \sum_{i=1}^{K}\bm{w}_{i}^{H}\mathbf{h}_{k}\mathbf{h}_{k}^H\bm{w}_{i}
			\!-\!\lambda\left(\sum_{k=1}^{K}\bm{w}_{k}^{H}\bm{w}_{k}\!-\!P_{T}\right),
		\end{aligned}
	\end{eqnarray}
	where $\lambda$ is the Lagrangian dual variable.
	Taking the partial derivative of $f_{2.1b}^{(L)}(\lambda, \mathbf{W},\pmb{\xi}^{op})$ with respect to $\bm{w}_{k}$, we obtain 
	\begin{eqnarray}
		\begin{aligned} \label{Der}
			&\frac{\partial f_{2.1b}^{(L)}(\lambda, \mathbf{W},\pmb{\xi}^{op})}{\partial \bm{w}_{k}}=\\
			&2\sqrt{\tilde{\alpha}_{k}}\mathbf{h}_{k}\xi_{k}^{op}\!-\!
			2\sum_{k=1}^{K}|\xi_{k}^{op}|^{2} \sum_{i=1}^{K}\mathbf{h}_{k}\mathbf{h}_{k}^H\bm{w}_{i}
			\!-\!2\lambda \mathbf{I}_{M} \bm{w}_{i}.	
		\end{aligned}
	\end{eqnarray}
	Then, by solving $\partial f_{2.1a}^{(L)}(\lambda, \mathbf{W},\pmb{\xi}^{op}) / \partial \bm{w}_{k}=0$, we can get the optimal expression of $\bm{w}_{k}^{op}$ in \eqref{ww}. For $\lambda^{op}$, taking the partial derivative of $f_{2.1b}^{(L)}(\lambda, \mathbf{W},\pmb{\xi}^{op})$ with respect to $\lambda$, we have
	\begin{eqnarray}
		\begin{aligned}
			&\frac{\partial f_{2.1b}^{(L)}(\lambda, \mathbf{W},\pmb{\xi}^{op})}{\partial \lambda}=-\sum_{k=1}^{K}\bm{w}_{k}^{H}\bm{w}_{k}\!+\!P_{T}.
		\end{aligned}
	\end{eqnarray}
	By solving the equation ${\partial f_{2.1b}^{(L)}(\lambda, \mathbf{W},\pmb{\xi}^{op})}/{\partial \lambda}=0$, i.e., $\sum_{k=1}^{K}||\bm{w}_{k}^{op}||^{2}=P_{T}$, we can obtain all the feasible solutions for $\lambda^{op}$. Among all the solutions, the smallest one is chosen since $\lambda\geq 0$, $\sum_{k=1}^{K}\bm{w}_{k}^{H}\bm{w}_{k}-P_{T} \leq 0$,  and the dual function of $f_{2.1b}^{(L)}(\lambda,\mathbf{W},\pmb{\xi}^{op})$, i.e., $\min \limits_{\lambda} \mathcal{G}(\lambda)	=\min \max\limits_{\mathbf{W}, \pmb{\xi}^{op}} f_{2.1b}^{(L)}(\lambda,\mathbf{W},\pmb{\xi}^{op})$ increases with $\lambda$. Among all the feasible solutions, the smallest one is selected since minimizing $ \mathcal{G}(\lambda)$ equals to maximize $ f_{2.1b}^{(L)}(\lambda,\mathbf{W},\pmb{\xi}^{op})$. Finally, we can get $\lambda^{op}$ in (\ref{lamm}).
	\hfill $\blacksquare$
\section{Proof of Proposition 3}
\label{Appendix_B}
The optimization problem is
	\begin{eqnarray}
\begin{aligned}
(P3.2b)\quad &\max_{\pmb{\phi}_l,\pmb{\varepsilon}} \quad 	f_{3.2b}(\pmb{\phi}_l,\pmb{\varepsilon}),\\
& ~\text{s.t.}  \quad \quad|\phi_{l,n}|=1, \quad n=1,2,...,N_l,
\end{aligned}
\end{eqnarray}
with
\begin{equation}
\begin{split}
\hspace{-2mm}
f_{3.2b}(\pmb{\phi}_l,\pmb{\varepsilon})\!\!=\!\!&\sum_{k=1}^{K}\sqrt{\tilde{\alpha}_k}
\left(\varepsilon_k^*(\bar{\pmb{\phi}_l}\bm{z}_{l,k}+c_{l,k})\!+\!\!(\bm{z}_{l,k}^H\bar{\pmb{\phi}_l}^H\!+\!c_{l,k}^*)\varepsilon_k\right)\\
&-\sum_{k=1}^{K}|\varepsilon_k|^2\left(\sum_{i=1}^{K}|\bar{\pmb{\phi}_l}\bm{z}_{l,i}+c_{l,i}|^2+\sigma_0^2\right).
\end{split}
\end{equation}
In $(P3.2b)$,
taking partial derivative of $f_{3.2b}(\pmb{\phi}_l,\pmb{\varepsilon})$ with respect to $\varepsilon_{k}$, we obtain 
\begin{eqnarray}
	\begin{aligned}
		\frac{\partial f_{3.2b}(\pmb{\phi}_l,\pmb{\varepsilon})}{\partial \varepsilon_{k}}=&
		2\sqrt{\tilde{\alpha}_{k}}(\bar{\pmb{\phi}_l}\bm{z}_{l,k}+c_{l,k})\\
		&-2\varepsilon_{k}
		\left(\sum_{i=1}^{K}|\bar{\pmb{\phi}_l}\bm{z}_{l,i}+c_{l,i}|^{2}+\sigma_{0}^{2}\right).
	\end{aligned}
\end{eqnarray}
By setting $\partial f_{3.2b}(\pmb{\phi}_l,\pmb{\varepsilon})/ \partial \varepsilon_{k}=0$, we can get the optimal $\varepsilon_{k}^{op}$ in \eqref{var} as
\begin{eqnarray}
	\begin{aligned}
		\varepsilon_{k}^{op}=\frac{\sqrt{\tilde{\alpha}_{k}}(\bar{\pmb{\phi}_l}\bm{z}_{l,k}+c_{l,k})}
		{\sum_{i=1}^{K}|\bar{\pmb{\phi}_l}\bm{z}_{l,i}+c_{l,i}|^{2}+\sigma_{0}^{2}}.
	\end{aligned}
\end{eqnarray}
Substituting $\varepsilon_{k}^{op}$ into $f_{3.2b}(\pmb{\phi}_l,\pmb{\varepsilon})$, we have  
\begin{eqnarray}
	\begin{aligned} \label{phi3}
	&f_{3.2b}(\pmb{\phi}_l,\pmb{\varepsilon}^{op})\\
			&=\sum_{k=1}^{K}\sqrt{\tilde{\alpha}_k}
		\left((\varepsilon_k^{op})^*(\bar{\pmb{\phi}_l}\bm{z}_{l,k}\!+\!c_{l,k})\!+\!(\bm{z}_{l,k}\bar{\pmb{\phi}_l}^H\!\!+\!\!c_{l,k}^*)\varepsilon_k^{op}\right)\\
		&\quad\quad-\sum_{k=1}^{K}|\varepsilon_k^{op}|^2\left(\sum_{i=1}^{K}|\bar{\pmb{\phi}_l}\bm{z}_{l,i}+c_{l,i}|^2+\sigma_0^2\right)\\
		&=-\bar{\pmb{\phi}_l}\mathbf{P}\bar{\pmb{\phi}_l}^H+2 \Re\lbrace\bar{\pmb{\phi}_l}\bm{d}_l+q_1\rbrace-c_1.
	\end{aligned}
\end{eqnarray}
where 
\begin{eqnarray}
\begin{aligned}
\mathbf{P}= \sum_{k=1}^K|\varepsilon_{k}^{op}|^2\sum_{i=1}^K\bm{z}_{l,i}\bm{z}_{l,i}^H,
	\end{aligned}
\end{eqnarray}
and
\begin{equation}
\begin{split}
\bm{d}_{l}=&
\sum_{k=1}^{K}\sqrt{\tilde{\alpha}_k}(\varepsilon_{k}^{op})^*\bm{z}_{l,k} 
- \sum_{k=1}^K|\varepsilon_k^{op}|^2\sum_{i=1}^{K}\bm{z}_{l,i}c_{l,i}^*.
\end{split}
\end{equation}
Among the denotations,
$q_1$ and $c_1$ are complex numbers, which are given by $q_1=\sum_{k=1}^K\sqrt{\tilde{\alpha}_k}(\varepsilon_{k}^{op})^*c_{l,k}$ and 
$c_1=\sum_{k=1}^{K}|\varepsilon_{k}^{op}|^2\sum_{i=1}^Kc_{l,i}c_{l,i}^*+\sum_{k=1}^K|\varepsilon_{k}^{op}|^2\sigma_0^2$, respectively.
In \eqref{phi3}, the first term $\bar{\pmb{\phi}_l}\mathbf{P}\bar{\pmb{\phi}_l}^H$ can be further expanded as
\begin{equation}
	\begin{split} \label{TR2}
		\bar{\pmb{\phi}_l}\mathbf{P}\bar{\pmb{\phi}_l}^H=&
		\sum_{i=1}^{(L-l+1)N_l}\sum_{j=1}^{(L-l+1)N_l}\bar{\phi}_{l,i}p_{i,j}\bar{\phi}^*_{l,j}\\
				=&\bar{\phi}_{l,n}p_{n,n}\bar{\phi}^*_{l,n}+\sum_{j=1,j\neq n}^{(L-l+1)N_l}\bar{\phi}_{l,n}p_{n,j}\bar{\phi}^*_{l,i}\\
				&+\!\!\!\!\!\!\!\sum_{i=1,i\neq n}^{(L-l+1)N_l}\!\!\!\!\!\bar{\phi}_{l,i}p_{i,n}\bar{\phi}^*_{l,n}
				\!\!+\!\!\!\!\!\!\sum_{i=1,i\neq n}^{(L-l+1)N_l}\sum_{j=1,j\neq n}^{(L-l+1)N_l}\!\!\!\!\!\!\bar{\phi}_{l,i}^*p_{i,j}\bar{\phi}_{l,j}\\
		\overset{(a)}{=}&
		\bar{\phi}_{l,n}p_{n,n}\bar{\phi}_{l,n}^*+2\Re \left\lbrace\sum_{j=1,j\neq n}^{(L-l+1)N_l}\bar{\phi}_{l,n}p_{n,j}\bar{\phi}_{l,i}^*\right\rbrace\\
		&+\sum_{i=1,i\neq n}^{(L-l+1)N_l}\sum_{j=1,j\neq n}^{(L-l+1)N_l}\bar{\phi}_{l,i}p_{i,j}\bar{\phi}^*_{l,j}.
	\end{split}
\end{equation}
In \eqref{TR2}, $(a)$ holds since $\mathbf{P}=\sum_{k=1}^K|\varepsilon_{k}^{op}|^2\sum_{i=1}^K\bm{z}_{l,i}\bm{z}_{l,i}^H$ is a Hermitian matrix and satisfies $p_{i,j}=p_{j,i}^*$.

The second term $\bar{\pmb{\phi}_l}\bm{d}_l$ of \eqref{phi3} can be further expanded as
\begin{equation}
	\begin{split}\label{TR1}
		\bar{\pmb{\phi}_l}\bm{d}_l
		=\sum_{i=1}^{(L-l+1)N_l}\bar{\phi}_{l,i}d_{l,i} 
		=\bar{\phi}_{l,n}d_{l,n}+\sum_{i=1,i\neq n}^{(L-l+1)N_l}\bar{\phi}_{l,i}d_{l,i}.
	\end{split}
\end{equation}

Substituting \eqref{TR1} and \eqref{TR2} into \eqref{phi3} and ignoring the constant term $c_1$, $f_{3.2b}(\pmb{\phi}_l,\pmb{\varepsilon}^{op})$ can be rewritten as
\begin{equation}
\begin{split} \label{phi23}
		f_{3.2c}(\bar{\phi}_{l,n},\pmb{\varepsilon}^{op})= &-\bar{\phi}_{l,n}p_{n,n}\bar{\phi}^*_{l,n}\\
		& +\!2\Re\left\lbrace \bar{\phi}_{l,n}(d_{l,n}\!-\!\!\!\!\!\sum_{j=1,j\neq n}^{(L-l+1)N_l}p_{n,j}\bar{\phi}^*_{l,j}) \right\rbrace\!+\!c_2\\
		\overset{(b)}{=}&-p_{n,n}
		+2\Re\lbrace \bar{\phi}_{l,n}\eta_n \rbrace+c_2,
	\end{split}
\end{equation}
where $(b)$ holds since $\bar{\phi}_{l,n}\bar{\phi}_{l,n}^*=1$ always holds and $\eta_n=d_{l,n}-\sum_{j=1,j\neq n}^{(L-l+1)N_l}p_{n,j}\bar{\phi}^*_{l,j}$, the constant term $c_2$ is represented as
\begin{equation}
\begin{split}
c_2=&-\sum_{i=1,i\neq n}^{(L-l+1)N_l}\sum_{j=1,j\neq n}^{(L-l+1)N_l}\bar{\phi}_{l,i}p_{i,j}\bar{\phi}^*_{l,j}\\
&+2\Re \lbrace 
\sum_{i=1,i\neq n}^{(L-l+1)N_l}\bar{\phi}_{l,i}d_{l,i}+q_1 \rbrace,
	\end{split}
\end{equation}
which is independent of $\bar{\phi}_{l,n}$.
 
From \eqref{phi23}, we obtain that maximizing $f_{3.2c}(\bar{\phi}_{l,n},\pmb{\varepsilon}^{op})$ is equivalent to maximizing $2\Re\lbrace \bar{\phi}_{l,n}\eta_n \rbrace$, which gives rise to the solution in \eqref{sphi}.
\hfill $\blacksquare$

\section{Convergence analysis of Algorithm 1}
\label{Appendix_D}
In Algorithm 1, Proposition \ref{alpha1} ensures that the update of $\pmb{\alpha}$ always produces an improving value, thus the objective function in \eqref{OBM1} always satisfies $f_{2}(\pmb{\alpha}^i,\mathbf{W}^{i-1},\pmb{\phi}_1^{i-1},...,\pmb{\phi}_L^{i-1}) \geq f_{2}(\pmb{\alpha}^{i-1},\mathbf{W}^{i-1},\pmb{\phi}_1^{i-1},...,\pmb{\phi}_L^{i-1}) $ in the $i$-th iteration.

For the BS's transmit beamforming optimization, when $\pmb{\alpha}$ and $\pmb{\phi}_1,...,\pmb{\phi}_L$ are fixed, the function in \eqref{15} is a biconvex problem with respect $\bm{w}_{k}$ and $\xi_{k}$. For fixed $\mathbf{W}$, the optimal $\xi_{k}^{op}$ can be obtained by setting $\partial f_{2.1b}(\mathbf{W},\pmb{\xi})/\partial \xi_{k}=0$.
For fixed $\xi_{k}$, the optimization problem to get the optimal $\mathbf{W}$ can be expressed as
\begin{eqnarray}
	\begin{aligned}
		\max_{\mathbf{W}} \quad f_{2.1b}(\mathbf{W})=&\sum_{k=1}^{K}\sqrt{\tilde{\alpha}_{k}}\left(\xi_{k}^{*}\mathbf{h}_{k}^{H}\bm{w}_{k}+\bm{w}_{k}^{H}\mathbf{h}_{k}\xi_{k}\right)\\
		&-\sum_{k=1}^{K}|\xi_{k}|^{2} \sum_{i=1}^{K}\bm{w}_{i}^{H}\mathbf{h}_{k}\mathbf{h}_{k}^{H}\bm{w}_{i},\\
		\text{s.t.}\quad\quad\quad\quad\quad\quad& \sum_{k=1}^{K}||\bm{w}_{k}||^{2}\leq  P_{T}.
	\end{aligned}
\end{eqnarray}	
The corresponding Lagrangian function is given by 
\begin{eqnarray} \label{erci}
	\begin{aligned}
		&f_{2.1b}^{(L)}(\mathbf{W})=-\sum_{k=1}^{K}|\xi_{k}|^{2}\sum_{i=1}^{K}\bm{w}_{i}^{H}\mathbf{h}_k\mathbf{h}_{k}^{H}\bm{w}_{i}\\
		&\!\!+\!\!\sum_{k=1}^{K}\sqrt{\tilde{\alpha}_{k}}\left(\xi_{k}^{*}\mathbf{h}_{k}^{H}\bm{w}_{k}\!\!+\!\!\bm{w}_{k}^{H}\mathbf{h}_{k}\xi_{k}\right)
		\!\!-\!\!\lambda \left(\sum_{k=1}^{K}\bm{w}_{k}^{H}\bm{w}_{k}\!\!-\!\!P_{T}\right).
	\end{aligned}
\end{eqnarray}
By setting $\partial f_{2.1b}^{(L)}(\mathbf{W}) / \partial \bm{w}_{k}=0$, the optimal expression $\bm{w}_{k}$ in \eqref{ww} can be obtained. Thus, $f_{2}(\pmb{\alpha}^i,\mathbf{W}^i,\pmb{\phi}_1^{i-1},...,\pmb{\phi}_L^{i-1}) \geq f_{3}(\pmb{\alpha}^{i-1},\mathbf{W}^{i-1},\pmb{\phi}_1^{i-1},...,\pmb{\phi}_L^{i-1}) $ is guaranteed.

With fixed $\mathbf{W}$, the objective functions
$f_{3.2b}(\pmb{\phi}_l,\pmb{\varepsilon})$ in \eqref{phi22} is solved by deriving the optimal solution for $\pmb{\phi}_l$, $l=1,2,...,L$ and $\pmb{\varepsilon}$, respectively. Specifically, we get the optimal $\varepsilon_{k}^{op}$ by setting $\partial f_{3.2b}(\pmb{\phi}_2,\pmb{\varepsilon})/ \partial \varepsilon_{k}=0$. Thus, we can optimize $\pmb{\phi}_l$ sequentially according to \eqref{sphi}. 
Based on the above analysis, we have $f_{3}(\pmb{\alpha}^i,\mathbf{W}^i,\pmb{\phi}_1^i,...,\pmb{\phi}_L^i) \geq f_{3}(\pmb{\alpha}^{i-1},\mathbf{W}^{i-1},\pmb{\phi}_1^{i-1},...,\pmb{\phi}_L^{i-1}) $ after the $i$-th iteration. Thus, the algorithm generates a non-decreasing sequence. 

To prove the convergence of Algorithm 1, we also need to show that  the sequence generated from Algorithm 1 is upper bounded. As shown in \eqref{AP}, for $ \alpha_k \geq 0$ and $\pmb{\alpha}=[\alpha_{1},\alpha_{2},...,\alpha_{K}]^{T}$, $f_{2}(\pmb{\alpha},\mathbf{W},\pmb{\phi}_1,...,\pmb{\phi}_L)$ is an increasing function with respect to $\pmb{\alpha}$. The maximum value of $f_{2}(\pmb{\alpha},\mathbf{W},\pmb{\phi}_1,...,\pmb{\phi}_L)$ with respect to $\pmb{\alpha}$ is achieved by setting $\alpha_k^{op} =\gamma_k$, leading to $f_{2}(\pmb{\alpha},\mathbf{W},\pmb{\phi}_1,...,\pmb{\phi}_L)=f(\mathbf{W},\pmb{\phi}_1,...,\pmb{\phi}_L)$. This shows that $f_{2}(\bm{\alpha},\mathbf{W},\pmb{\phi}_1,...,\pmb{\phi}_L)$ is upper-bounded by the original objective function $f(\mathbf{W},\pmb{\phi}_1,...,\pmb{\phi}_L)$. Since $\gamma_k$ ranges from $0$ to $\gamma_k^{max}$, $f(\mathbf{W},\pmb{\phi}_1,...,\pmb{\phi}_L)$ must have a finite upper bound.  Thus, Algorithm 1 generates a bounded non-decreasing sequence and hence converges. 
\hfill $\blacksquare$
	\bibliographystyle{IEEEtran}
	\bibliography{Rreference}
\end{document}